\documentclass[english,preprint,12pt]{elsarticle}
\usepackage[LGR,T1]{fontenc}
\usepackage[latin9]{inputenc}
\usepackage{color}
\usepackage{array}
\usepackage{float}
\usepackage{calc}
\usepackage{textcomp}
\usepackage{amsmath}
\usepackage{graphicx}
\usepackage{setspace}
\usepackage{nomencl}
\providecommand{\printnomenclature}{\printglossary}
\providecommand{\makenomenclature}{\makeglossary}
\makenomenclature

\makeatletter

\DeclareRobustCommand{\greektext}{%
  \fontencoding{LGR}\selectfont\def\encodingdefault{LGR}}
\DeclareRobustCommand{\textgreek}[1]{\leavevmode{\greektext #1}}
\DeclareFontEncoding{LGR}{}{}
\DeclareTextSymbol{\~}{LGR}{126}
\providecommand{\tabularnewline}{\\}

\usepackage{graphicx}
\usepackage{natbib}
\usepackage{txfonts}
\usepackage{amssymb}
\usepackage{fleqn}
\usepackage{multicol}
\biboptions{sort&compress}
\usepackage{breqn}

\gdef\wrap@breqn@environ#1#2{
\expandafter\let\csname breqn@oldbegin@#1\expandafter\endcsname\csname #1\endcsname
\expandafter\let\csname breqn@oldend@#1\expandafter\endcsname\csname end#1\endcsname
\expandafter\gdef\csname breqn@begin@#1\endcsname{%
\expandafter\let\csname #1\expandafter\endcsname\csname breqn@oldbegin@#1\endcsname%
\begin{#2}%
}
\expandafter\gdef\csname breqn@end@#1\endcsname{%
\expandafter\let\csname end#1\expandafter\endcsname\csname breqn@oldend@#1\endcsname%
\end{#2}%
\expandafter\let\csname #1\expandafter\endcsname\csname breqn@begin@#1\endcsname%
\expandafter\let\csname end#1\expandafter\endcsname\csname breqn@end@#1\endcsname%
}
\expandafter\let\csname #1\expandafter\endcsname\csname breqn@begin@#1\endcsname
\expandafter\let\csname end#1\expandafter\endcsname\csname breqn@end@#1\endcsname
}
\wrap@breqn@environ{equation}{dmath}
\wrap@breqn@environ{equation*}{dmath*}
\usepackage{hyperref}

\usepackage{float}
\floatstyle{plaintop}
\restylefloat{table}
\usepackage[labelfont=bf,justification=raggedright,singlelinecheck=false]{caption}
\makeatother

\usepackage{babel}
\begin{document}
\begin{frontmatter}

\title{A methodology for using Kalman filter to determine material parameters
from uncertain measurements}

\author[rvt,focal]{Abdallah Shokry\corref{cor1}\fnref{fn1}}

\ead{abdallah.shokry@fayoum.edu.eg}

\author[rvt]{Per Ståhle}

\ead{per.stahle@solid.lth.se}

\cortext[cor1]{Corresponding author}

\fntext[fn1]{+46 73 692 5606}

\address[rvt]{Division of Solid Mechanics, Lund University, 22100 Lund, Sweden}

\address[focal]{Industrial Engineering Department, Fayoum University, 63514 Fayoum,
Egypt}
\begin{abstract}
A Kalman filter can be used to determine material parameters using
uncertain experimental data. However, starting with inappropriate
initial values for material parameters might include false local attractors
or even divergence. Also, inappropriate choices of covariance errors
of initial state, present state, and measurements might affect the
stability of the prediction. The present method suggests a simple
way to predict the parameters and the errors, required to start Kalman
filter based on known parameters that are used to generate the data
with different noises used as ``measurement data''. The method consists
of two steps. First, an appropriate range of parameter values is chosen
based on a graphical representation of the mean square error. Second,
the Kalman filter is used based on the selected range and the suggested
parameters and errors. The method of the filter significantly reduces
the iteration time, and covers a wide range of initial suggested values
for the parameters compared with the standard Kalman filter. When
the methodology is applied to real data, very good results are obtained.
Diffusion coefficient for bovine bone is chosen to be a case study
in this work.\end{abstract}
\end{frontmatter}

\section{Introduction}

The Kalman filter is an inverse method to determine variables or parameters
using input data with more noise and get output data with less noise.
It is firstly presented by R.E. Kalman \citep{Kalman(1960)} in 1960.
Kalman filter has the advantages of taking the random noise for state
and measurements into consideration, also it is an optimal estimator
for linear models because it minimizes the mean square error between
the state. In addition, it converges quickly. A more complete introduction
to the Kalman filter is given by Brown \citep{Brown(1983)}. The Kalman
filter can be found under different updated forms that used in many
different fields such as tracking objects \citep{Siouris(1997),Weng(2006),Antnov(2011)},
control systems \citep{Ahn(2009),Shi(2009)}, and weather forecast
\citep{Mitchell(2009),Wu(2010),Miyoshi(2012)}.

\begin{singlespace}
\noindent %
\fbox{\begin{minipage}[t]{1\columnwidth}%
\begin{spacing}{0}
\noindent  \printnomenclature[3.5em]\end{spacing}
\end{minipage}}
\end{singlespace}

Kalman filter can be used to determine material parameters from uncertain
and inaccurate measurements. Aoki et al. \citep{Aoki(1997)} used
Kalman filter to identify Gurson's model constant. They found that
the accuracy of parameters prediction is affected by both specimen
geometry and measurement type, and the shape of the tested specimen
affects the convergence of the parameters. Also, they noticed that
the rate of convergence can be improved by combining measurements
of two different specimens in shape. The identification of Gurson\textendash Tvergaard
material model parameters via Kalman filtering technique is studied
by Corigliano et al. \citep{Corigliano(2000)}. They stated that the
estimated values of the parameters are in well agreement with those
obtained in previous work, but the initial suggested values for the
seeking parameters affects the estimated parameters. 

Nakamura et al. \citep{Nakamura(2007)} implemented Kalman filter
to determine elastic-plastic anisotropic parameters for thin materials
using instrumented indentation. They observed that the initial chosen
values for the parameters converged to a specific small area, but
not to one point. Also, based on the convergence intensity, the parameters
are determined. The same findings are obtained by using Kalman filter
to determine the nonlinear properties of thermal sprayed ceramic coatings
\citep{Nakamura(2007)2}. Bolzon et al. \citep{Bolzon(2002)} used
Kalman filter to identify parameters of a cohesive crack model. They
reported that almost a linear correlation between convergent parameters
is found, and the reason for the multiple local minimum might be related
to using the linear Kalman filter for non-linear model. 

Vaddadi et al. \citep{Vaddadi(2003)} used Kalman filter to determine
critical moisture diffusion parameters for a fiber reinforced composite.
They estimated the parameters from the intensity of the convergence,
which found to be in consistent with known values. Another study made
by Vaddadi et al. \citep{Vaddadi(2007)} to determine hygrothermal
properties in fiber reinforced composite using Kalman filter. The
parameters are extracted by reading the intensity of convergence plot. 

Kalman filter is an efficient way to filter noisy experimental data
for determination of material parameters. However, the initial suggested
parameters required for Kalman filter should be chosen carefully,
to avoid false local attractor. Also, the covariance error for the
parameters noise almost assumed to be zero, which slow the rate of
convergence and might lead to more than one intensity area for the
predicted parameters. 

In this study, a methodology will be applied for using Kalman filter
to determine material parameters using uncertain measurements. The
methodology starts by a way based on the mean square error to choose
appropriate initial suggested parameters required for Kalman filter,
and followed by a suggested way to choose the covariance errors for
both state and measurements. The determination of diffusion coefficients
in bovine bone for generated data with different noises scatter from
known parameters will be applied as a case study. A real measurements
will be used also.

\section{Methods}

\subsection{The Model}

Assume that an experiment resulted in $N$ measurements obtained at
different times, locations, temperatures etc. These are collected
in a vector, $z$\nomenclature{$z$, $z_{k}$, $z_{av}$}{vectors with $N$ measurements, at iteration $k$, avergare $z$ },
with $N$\nomenclature{$N$, $n$, $M$}{number of measurements, state variables, iterations}
measurements. The experimental data may be obtained at different known
times, locations, temperatures etc. Measurements and all other data
are available \textit{a priori}.

In an attempt to predict the measurements a model, $h=h(x)$\nomenclature{$h$}{vector of $N$ predicted measurements}
is used, with $h$ being a vector of $N$ predictions of observations.
Further, $x$\nomenclature{$x$, $\bar{x}$, $x_{0}$}{vectors of $n$ unknown, perturbed, initial state variables}
is a vector of $n$ unknown parameters defining the model based on
variables such as position, temperature, time, etc. The unknown model
parameters may describe the state of the system regarding, material,
geometry or similar. In the present study, $x$ is limited to parameters
describing the material.

Measurements always include systematic and non-systematic errors due
to instrumentation, indirect observations, gauges sensitive, irrelevant
external influence, and similar. Material parameter is sought but
the experimental method may require a state parameters to be determined
as well. Further, material parameters contain non-systematic errors
due to thermal fluctuations, unstable structural configurations such
as mobile dislocations, impurities, inclusions, unstable chemical
composition, etc. Also inevitably, there is a difference between model
and reality while a model never gives an exact description of the
physical processes. Under ideal conditions the model would perfect
in the sense that $z=h(x)$. Here, only non-systematic errors or noise
is considered. The model is defined for measurement $i$ as

\begin{equation}
z=h(\bar{x})+v\,,\label{eq:model-properties}
\end{equation}
where $v$\nomenclature{$\nu$, $\nu_{k}$}{vectors of $N$ errors of the measurements, at iteration $k$}
is a vector with $N$ errors due to inaccurate measurements $z$.
The instant parameter $\bar{x}$ corresponding to the individual measurement
$i$ includes noise according to

\begin{equation}
\bar{x}\mbox{=}x+w\,,\label{eq:matparameter noise}
\end{equation}
where $w$\nomenclature{$w$, $w_{k}$}{vectors of $n$ perturbations of the state variables, at iteration $k$}
is a vector with $n$ errors caused by the parameter deviations. The
elements of $v$ and $w$ are assumed to be uncorrelated. All elements
of $w$ and $v$ are supposed to be random, having the same respective
stochastic distribution and for both a vanishing mean value is expected,
cf. \citep{Brown(1983)}.

Assuming that a set of parameters $\hat{x}_{k}$\nomenclature{$\hat{x}_{k}$}{vector of  estimated state variables at iteration $k$}
is an estimate in the neighborhood of $x$, an improved estimate $x_{k+1}$
may be obtained by using linearized using a Taylor series which gives
\begin{equation}
h(\hat{x}_{k+1})\text{\ensuremath{\approx}}h(\hat{x}_{k})+\mathbf{H}(\hat{x}_{k})(x_{o}-\hat{x}_{k})\,,\label{eq:Taylor}
\end{equation}
when quadratic and higher order terms of $x$ are neglected. On matrix
form involved variables are

\begin{equation}
h(x)\mbox{=}\begin{bmatrix}h^{(1)}\\
\vdots\\
h^{(N)}
\end{bmatrix},\mathbf{\ H}(x)\mbox{=}\begin{bmatrix}\dfrac{\partial h^{(1)}}{\partial x_{1}} & \cdots & \dfrac{\partial h^{(1)}}{\partial x_{n}}\\
\vdots & \ddots & \vdots\\
\dfrac{\partial h^{(N)}}{\partial x_{1}} & \cdots & \dfrac{\partial h^{(N)}}{\partial x_{n}}
\end{bmatrix},\ x=\begin{bmatrix}x_{o}\\
\vdots\\
x_{n}
\end{bmatrix}\,.
\end{equation}
Here, $\mathbf{H}$\nomenclature{$\mathbf{H}$, $\mathbf{H}_{k}$}{matrix of derivatatives of $h$, at iteration $k$}
is an $N\times n$ a Jacobian matrix. \nomenclature{$T$}{transpose}

\subsection{Least-Squares}

The system is supposed to be overdetermined, meaning that the number
of measurements $N$ exceeds the number of unknown parameters $n$.
\textcolor{black}{The best fit in the least-squares sense minimizes
the sum of squared residuals with a residual being the difference
between an observations }$z$ \textcolor{black}{and the predictions
}$h(x)$.\textcolor{black}{{} The unknown }$x$\textcolor{black}{{} is
obtained in a series of iterative improvements of the approximation
}$x\approx\hat{x}_{k}$\textcolor{black}{, where }$k$\textcolor{black}{{}
is the iteration number. The initial parameters }$x_{0}$\textcolor{black}{{}
may be an educated guess based on previous measurements, data from
resembling materials or other similar expectatio}ns.

Solutions for non-linear systems (see \ref{sec:Appendix A}) may be
obtained iteratively. As an example, the Newton-Raphson method applied
to these solutions gives the following recursive scheme,

\begin{equation}
\hat{x}_{k+1}=\hat{x}_{k}+(\mathbf{H}_{k}^{\text{T}}\mathbf{H}_{k})^{-1}\mathbf{H}_{k}^{\text{T}}\{z-h(\hat{x}_{k})\},\label{eq:iterated least square}
\end{equation}
where $\mathbf{H}_{k}$ is an $N\times n$ matrix and a function of
$\hat{x}_{k}$. If convergence is reached, a local minimum of the
sum of squared residuals has been found. To find global minimum, additional
steps have to be taken. The drawback of the method is that convergence
is not necessarily reached and is less likely if the \textit{a priori}
information of $x$ is vague, unreliable or even misleading. This
is especially accentuated when the measurements are noisy. Further,
all data have to be present \textit{a priori}. Modifications have
been developed that allow an incremental treatment, which may be useful
if data is continuously added (cf. \citep{Plackett(1950)}).

\subsection{Kalman Filter\label{sub:Kalman-Filter}}

The method of least-squares does not take the properties of the noises
$v$ and $w$ as regards expected mean value and distribution into
account. Opposed to that, the Kalman filter is developed to use information
about the noise \textit{a priori} or as the measurements are assembled.
The filter is an improvement of the least-squares method as it recursively
optimizes the unknown model parameters in search for least sum of
squared errors. The method is developed with the particular endeavor
to effectively handle noisy input data \citep{Brown(1983)}. It is
also operating incrementally so that new may be added as they appear
in during ongoing measurements without loss of accuracy. However,
in the present study all measurements are supposed to be available
when the optimization is initiated. The Kalman method is generally
assumed to be an effective method to filter noisy data combined with
a high convergence rate. The derived algorithm is taking the character
of the noise into consideration. The expected vanishing mean values
for $\nu$ and $w$ are explicitly utilized.

Assuming that the measured data is the predicted data based on the
optimum material parameters with the addition of noise, as given by
Eq. (\ref{eq:model-properties}) it is here assumed that the noise
$\nu$ a distribution with a zero mean value.

Initially a set of parameters $x_{0}$ is selected based on \textit{a
priori} information from other measurements under same or similar
conditions or otherwise known data. From this, the parameters are
iteratively updated using an algorithm on the same form as the least-square
algorithm (cf. Eq. \ref{eq:iterated least square}) as follows

\begin{equation}
\hat{x}_{k+1}=\hat{x}_{k}+\mathbf{K}{}_{k}\{z_{k}-h(\hat{x}_{k})\}\ \ \ \text{for}\ \ \ k\mbox{=}0,1,2,...M\,,\label{eq:kalman}
\end{equation}
where $\mathbf{K}{}_{k}$\nomenclature{$\mathbf{K}{}_{k}$}{$n\times N$ matrix represents Kalman gain}
is an $n\times N$ matrix denoted the Kalman gain, and $M$ is the
number of iterations. In Eq. (\ref{eq:kalman}). Normally the Kalman
algorithm operates on single measurements one by one so that the Kalman
gain is updated for every new measurements. This may be necessary
for interactive processes where the action requires knowledge of the
instantaneous state of the system. In the present study, iterations
are simultaneously utilizing all measurements. The derivation of the
Kalman gain (see \ref{sec:Appendix-B}) is based on measurements added
recursively as in the original form of the filter.

The optimal $\mathbf{K}{}_{k}$ that minimizes the mean square error
is given by 

\begin{equation}
\mathbf{K}{}_{k}=\boldsymbol{P}_{k}^{-}\mathbf{H}_{k}^{\text{T}}(\mathbf{H}_{k}\boldsymbol{P}_{k}^{-}\mathbf{H}_{k}^{\text{T}}+\boldsymbol{R}_{k})^{-1}\,.\label{eq:Kalman gain-2}
\end{equation}
where $\boldsymbol{P}_{k}^{-}$\nomenclature{$\boldsymbol{P}_{k}^{-}$, $\boldsymbol{P}_{k}$}{variance errors of the parameters before and after iterative update at iteration $k$}
is the variance of the errors of the parameters before the iterative
update, and $R_{k}$\nomenclature{$\boldsymbol{R}_{k}$, $\boldsymbol{R}^{(i)}$}{covariance error for measurements at iteration $k$, constant covariance at initial parameters $i$}
is $N\times N$ matrix introduces the covariance error of the measurements
that computed as 

\begin{equation}
\boldsymbol{R}_{k}=\text{Var}(v_{k})\,.\label{eq:Covariance error for measurements}
\end{equation}

The $\boldsymbol{P}_{k+1}^{-}$ is expressed as 

\begin{equation}
\boldsymbol{P}_{k+1}^{-}=(\boldsymbol{I}-\mathbf{K}{}_{k}\mathbf{H}_{k})\boldsymbol{P}_{k}^{-}+\boldsymbol{Q}_{k}\,,\label{eq:P-recursion-2}
\end{equation}
where $\boldsymbol{Q}_{k}$ is an $n\times n$ matrix represents the
covariance errors for the state parameters that is computed as

\begin{equation}
\boldsymbol{Q}_{k}=2\text{Var}(w_{k}),\label{eq:def_of_Q_k}
\end{equation}

The derivation of $\mathbf{K}{}_{k}$, $\boldsymbol{P}_{k}^{-}$,
$\boldsymbol{R}_{k}$, and $\boldsymbol{Q}_{k}$ are given in details
in \ref{sec:Appendix-B}. 

The algorithm involves recursive use of the Eqs. (\ref{eq:kalman}),
(\ref{eq:Kalman gain-2}) and (\ref{eq:P-recursion-2}). The measurements
considered in each recursive cycle may be everything from a single
measurement to all measurements. For non-linear problems, each cycle
may be repeated until convergent results are obtained. As an alternative,
the entire recursive scheme may be restarted and the resulting parameters
$x_{n}$, $\boldsymbol{P}_{n}^{-}$, from previous application of
the scheme are used as initial parameters. Also $\boldsymbol{Q}_{k}$
and $\boldsymbol{R}_{k}$ may be adjusted based on the improved information
that is obtained. In the present study, all measurements are placed
in a single set with $N$ measurements, meaning that $z_{k}=z$ is
a constant vector with $N$ elements. The number of recursive cycles
is $M$ and $k=1,2,...,M$ where $M$ is prescribed or conditional.
The recycling is performed to achieve a converged result for a non-linear
problem. In each cycle, $x_{k}$, $\mathbf{H}_{k}$, $\boldsymbol{P}_{k}$
and therefore $\mathbf{K}{}_{k}$ is updated. 

By putting $\boldsymbol{R}_{k}=R\boldsymbol{I}$ in Eq. (\ref{eq:Kalman gain-2})
a\nomenclature{$A$, $B$, $R$}{three constants}nd then taking the
limiting result as $R\rightarrow0$ one obtains $\mathbf{K}{}_{k}=\mathbf{H}_{k}^{\text{T}}(\mathbf{H}_{k}\mathbf{H}_{k}^{\text{T}})^{-1}$
$=(\mathbf{H}_{k}^{\text{T}}\mathbf{H}_{k})^{-1}\mathbf{H}_{k}^{\text{T}}$.
After inserting this into Eq. (\ref{eq:kalman}) it is readily seen
that the result is identical to that of the non-linear least square
method, cf. Eq. (\ref{eq:iterated least square}). The result is independent
of $\boldsymbol{P}_{k}^{-}$ and consequently also independent of
$\boldsymbol{Q}_{k}$.

\section{Methodology}

In \citep{Aoki(1997),Nakamura(2007),Vaddadi(2007)}, the $\boldsymbol{R}_{k}$
was chosen to be a small percentage of the measured data, and the
$\boldsymbol{Q}_{k}$ value was chosen to be zero. In a recent study,
the $\boldsymbol{R}_{k}$ value was chosen as the difference between
the measured data and a predicted data a round the measured data,
and the $\boldsymbol{Q}_{k}$ was chosen to be unity \citep{Lindberg(2014)}.

To use the Kalman filter the parameters $\boldsymbol{Q}_{k}$ and
$\boldsymbol{R}_{k}$, and initial values for $\hat{x}_{0}$ and $\boldsymbol{P}_{0}^{-}$
have to be defined. In the following, different strategies for choosing
these values is described.

For common usage of the Kalman filter, the choice would be $\boldsymbol{R}_{k}=\text{Var}(\nu_{k})$
and $\boldsymbol{Q}_{k}=2\text{Var}(w_{k})$ according to Eqs. (\ref{eq:Covariance error for measurements})
and (\ref{eq:def_of_Q_k}). These variations are assumed to be known
\textit{a priori}, at least approximately. The information may be
based on expectation or derived from the present measurements, using
a large variety of hypotheses. When the method is used recursively,
the indices k allow for using $\boldsymbol{Q}_{k}$ or $\boldsymbol{R}_{k}$
or both as functions of time, position, etc.

The Kalman gain, given by Eq. (\ref{eq:Kalman gain-2}), with the
selected $\boldsymbol{Q}_{k}$ or $\boldsymbol{R}_{k}$, minimizes
the squared error of the estimate $\hat{x}_{k}$ of $x$. A condition
for the derivation is that $h(x)$ is a linear function of the parameters
$x$. In the present study, $h(x)$ is a non-linear function of $x$.
The aim is to formulate a strategy for selecting the free parameters
$\boldsymbol{R}_{k}$ and $\boldsymbol{Q}_{k}$, not necessarily according
to Eqs. (\ref{eq:Covariance error for measurements}) and (\ref{eq:def_of_Q_k}),
so that the square of the error

\begin{equation}
\mathcal{M}(x)=\frac{1}{N}\text{Var}(z-h(x))\,,\label{eq:def_square_error}
\end{equation}
is minimized. The study does not attempt to be exhaustive and the
conclusions are empirical and based on a case study. The selection
of method is primarily based on convergence rate. With a wide range
of starting values, occasionally the estimate converges outside the
range of interest and there is also the risk of failure in producing
converging results at all. These, disadvantages are also considered
in the selection of a suitable procedure for selecting $\boldsymbol{R}_{k}$
and $\boldsymbol{Q}_{k}$.

In the first part of the study, the initial parameters $\hat{x}_{0}$
are selected to cover a several orders of magnitude wide range of
values. Under normal circumstances, this cannot be done for non-linear
phenomena or realistic geometries or anything else for which an analytical
solution cannot be found, which may be the general case. When the
predictions are based on non-linear numerical calculations of field
problems, e.g., using lengthy finite element analyses, usually only
a few initial parameters $\hat{x}_{0}$ can be considered. Here however,
a wide range of initial parameters is examined as regards the mean
square error $\mathcal{M}(\hat{x}_{0})$\nomenclature{$\mathcal{M}(\hat{x})$}{mean square error between $z$ and $h$}.
The aim is to obtain an overall picture of the possibilities of fast
convergence or difficulties because of present local minima, sadle
points etc.

The second part is, the using of a suggested method and comparing
the resulting convergence rate, the ability of producing convergent
results, and the percentage of convergent results from a range of
the initial parameters with three additional methods. All four methods
are described in \autoref{sub:Selecting--and}.

\subsection{{\normalsize{}\label{sub:Selecting--and}Selecting }{\boldmath{$Q$}}$_{k}${\normalsize{}
and }{\boldmath{$R$}}}

In the present study, the results of constant 

\begin{equation}
{\boldsymbol{Q}_{k}=2p^{2}xx^{\text{T}}\quad\text{including}\quad\boldsymbol{Q}_{k}=0,}\label{eq:Q-1}
\end{equation}
are evaluated. The parameter $p$\nomenclature{$p$}{a priori estimated relative variance}
is an \textit{a priori} estimated relative variance.

The relative variance $p$ of the state parameters is a quantity that
possibly can be guessed with more or less accuracy. Since the real
state parameters $x$ are not known \textit{a priori} constant $\boldsymbol{Q}_{k}$
may be either too big or too small depending on how accurate initial
guess of $x_{0}$ is. However, since the knowledge of the state variables
increases as the iterations proceed, to stick a constant $\boldsymbol{Q}_{k}$
may unnecessarily slow down the convergence rate. As the iterations
proceed, the estimate $x_{k}$ is improved and a better estimate for
$\boldsymbol{Q}_{k}$ can be used. Here, an updated $\boldsymbol{Q}{}_{k}$
is also evaluated. With the variation of $x$ being $w_{k}=px$, the
unknown $x$ is here assumed to best approximated with $x_{k}$. The
updated value given by Eq. (\ref{eq:def_of_Q_k}) then becomes

\begin{equation}
\boldsymbol{Q}_{k}=2p^{2}\text{Var}(x_{k})\,.\label{eq:updated value of Q}
\end{equation}
The $\boldsymbol{R}_{k}$ taken as 

\begin{equation}
\boldsymbol{R}^{(i)}=v^{(i)}I,\,\, i\nolinebreak=1,2,....1681,\label{eq:noise max-1}
\end{equation}
which is assumed to be a reasonable approximation of Eq. (\ref{eq:Covariance error for measurements}),
where measurement noise $v^{(i)}$ is given by 

\begin{equation}
v^{(i)}=\max\vert\text{Var(}z_{N}-h_{N}(\hat{x}_{0}^{(i)}))\vert]\label{eq:noise max}
\end{equation}

where max denoted to the maximum value, and $i$ is the number of
initial parameters.

In the suggested method, the $\boldsymbol{Q}{}_{k}$ and $\boldsymbol{R}^{(i)}$
are chosen for large values based on the evaluation according to Eqs.
(\ref{eq:Kalman gain-2}) and (\ref{eq:P-recursion-2}), cf. \autoref{sec:Results-and-Discussions}
for explanation. There are additional two Kalman filter methods, one
method uses $p=0$ in Eq. (\ref{eq:Q-1}), and the another method
uses $p=0.01$ that gives the smallest standard deviation and largest
percentage of convergence among different values for $p$, the mean
values are closest to each other (see \autoref{fig:figure4}). These
two methods use $\boldsymbol{R}_{k}$ as 

\begin{equation}
\boldsymbol{R}=[\max\vert\text{Var(}z_{N}-h_{N}(x))\vert]I\,\,,\label{eq:noise measur-1}
\end{equation}

so, the $\boldsymbol{R}$ is the largest squared element for the variance
of the difference between measured data, $z$, and predicted data
with noise. Here, $\boldsymbol{Q}$ and $\boldsymbol{R}$ without
$k$ index denote constant covariances during iterations. 

The fourth method is the non-linear least square method, which is
obtained by letting $\boldsymbol{R}_{k}\rightarrow0$ as it is described
in \autoref{sub:Kalman-Filter}.

\subsection{{\normalsize{}The Initial Predicted Parameters Error }{\boldmath{$P$}}$_{0}^{-}$}

The initial predicted parameters error $\boldsymbol{P}_{0}^{-}$ is
an $n\times n$ matrix that contains the expected values for the errors
between the unknown parameters $x_{k}$ and the initial predicted
parameters $\hat{x}_{0}$ before the first iteration (see \ref{sec:Appendix-B}).
$\boldsymbol{P}_{0}^{-}$ is chosen as

\begin{equation}
\boldsymbol{P}_{0}^{-}=\text{Var}(\hat{x}_{0}^{(max)}-\hat{x}_{0}^{(min)})\,.\label{eq:pred-param-error}
\end{equation}
where $x_{0}^{(max)}$ and $x_{0}^{(min)}$ are the two vectors that
contain maximum and minimum of the selected initial values for the
parameters. The choice for $\boldsymbol{P}_{0}^{-}$ is the same in
all cases in the present study. Only one large initial value is tested
since the $\boldsymbol{P}_{0}^{-}$ introduces the variance between
the seeking parameters and the initial parameters.

\subsection{{\normalsize{}Summary of Methods}}

\begin{center}
\begin{table}[H]
\begin{centering}
\resizebox{\textwidth}{!}{
\begin{tabular}{>{\centering}m{0.3\textwidth}>{\centering}m{0.3\textwidth}>{\centering}m{0.2\textwidth}>{\centering}m{0.22\textwidth}>{\centering}m{0.1\textwidth}}
\hline 
Method & Q & R  & P  & D.N\%\tabularnewline
\hline 
Kalman filter 1 (suggested method) & $\boldsymbol{Q}=\boldsymbol{P}_{0}$ as in Eq. (\ref{eq:Initial covaraince error})  & Eq. (\ref{eq:covariance measurements error D B RLQL}) & Eq. (\ref{eq:pred-param-error}) & 92.09\tabularnewline
Kalman filter 2 & Eq. (\ref{eq:Q-1}) with $p=0$ & Eq. (\ref{eq:noise measur-1}) & Eq. (\ref{eq:pred-param-error}) & 80.90\tabularnewline
Kalman filter 3 & Eq. (\ref{eq:Q-1}) with $p=0.01$ & Eq. (\ref{eq:noise measur-1}) & Eq. (\ref{eq:pred-param-error}) & 80.19\tabularnewline
Non-linear least square & \_\_\_ & 0 & \_\_\_ & 40.93\tabularnewline
\hline 
\end{tabular}}
\par\end{centering}

\protect\caption{Methods used}
\label{Table:1}
\end{table}

\par\end{center}

\section{Case Studies}

The determination of the diffusion coefficient of mammal bone using
uncertain data is chosen to be a case study. Diffusion has recently
been suggested to play an important role in transporting substances
from the inner boundaries to the outer boundaries of bone. Therefore,
knowing the diffusion coefficients in human bone are important to
give required information for design of individual exercise programs
that maximizes bone remodeling and bone healing. 

In the present study, the proposed four methods are applied to several
simulated cases of generated data and the most effective method is
applied to a case of real experimental data. The real experiment measures
the amount of ions that leaves bovine bone samples that were put into
a container with distilled water. During elapsing time the conductivity,
$\zeta(t)$, of the water increases in proportion to the ionic concentration.
The experiment is reported in Lindberg et al. \citep{Lindberg(2014)}.
Cubic bone samples with the side length $L=$10.1\nomenclature{$L$}{side length of bone sample }
mm from a bovine long bone were used. The concentration were measured
using a SevenEasy S30 conductivity meter from Mettler Toledo with
an accuracy of \textpm 0.5\% of the measured value. Further details
regarding the experiment is found in \citep{Lindberg(2014)}.

The following model is suggested for the conductivity as a function
of time, $t$, \nomenclature{$t$}{time}

\begin{equation}
\zeta(t)=A-B\sum_{m=1}^{M}\frac{8}{\pi^{2}(2m-1)^{2}}\exp\{-(2m-1)^{2}\pi^{2}D\frac{t}{L^{2}}\}\:,\label{eq:pre-Conductivity equation}
\end{equation}
where $D$\nomenclature{$D$}{diffusion parameter} is the diffusion
constant, and $A$ and $B$ are unknown constants. The model is based
on Fick's law using relevant boundary conditions.

The constants $A$ and $B$ provide the relation between the concentration
in the bone sample and the conductivity in the distilled water. Putting
$t=0$ gives

\begin{equation}
A=\zeta(0)+B\:,
\end{equation}
where $\zeta(0)$ \nomenclature{$\zeta(t)$}{conductivity of escaped ions at time $t$}taken
to be the conductivity measured at $t=0$. By using $h(t)=\zeta(t)-\zeta(0)$
as the measured quantity, the following model is obtained,

\begin{equation}
h(t)=B(1-\sum_{m=1}^{M}\frac{8}{\pi^{2}(2m-1)^{2}}\exp\{-(2m-1)^{2}\pi^{2}D\frac{t}{L^{2}}\})\:,\label{eq:Conductivity equation}
\end{equation}
The two remaining constants $D$ and $B$ is determined using the
Kalman filter. More details are found in Lindberg et al. \citep{Lindberg(2014)}.

To the experiment is added a Monte Carlo set of 250 fictive measurements
where generated data is used as measurements. The measured data is
generated using the exact model Eqs. (\ref{eq:model-properties})
and (\ref{eq:Conductivity equation}) to compute the measurement vector
$z$. To provide realistic conditions, a variation is added to the
state parameters $D$ and $B$ and to the measurements. To this end,
a random noise of 5\%, i.e. $\left|w_{D}\right|<0.05D$ and $\left|w_{B}\right|<0.05B$,
is added. In the same way, noises $\nu$ of 10\%, 50\% and 100\% are
included in the generated measurements $z$, i.e. $z=h(t)+\nu$, where
$\nu=q\frac{1}{N}\sum h(t_{i})$ and $q$ equals 0.1, 0.5 and 1. Summation
is performed for the 24 different times of measurement. Further, the
$w_{D}$, $w_{B}$ and $\nu$ are uncorrelated and the probability
density is constant within the limits of the respective noise. Also
the noises for each measurements are uncorrelated. Thus, 

\begin{equation}
\text{Cov(}\vartheta_{i},\vartheta_{j})=\left\{ \begin{array}{ccc}
\vartheta_{i}^{2} & \text{if} & i=j\\
0 & \text{if} & i\neq j
\end{array}\right.\label{eq:cov definition}
\end{equation}
where $\vartheta_{i}$ and $\vartheta_{j}$ represents \nomenclature{$\vartheta_{i}$}{elements of noise vectors $w$, and $v$}
the elements of each noise \nomenclature{$w_{D}$, $w_{B}$}{noise of parameters $D$,  and $B$}
vector $w_{D}$, $w_{B}$ and $\nu$ at individual measurements $i$
and $j$. 

As observed from Eq. (\ref{eq:Conductivity equation}), the only available
time unit is provided by $L^{2}/D$. A typical experiment lasts for
around 11.7 time units, i.e., $0\leq t\leq11.7L^{2}/D$ with generated
measurements taken in time intervals of around $0.0032L^{2}/D$. 

In an attempt to obtain less non-linear formulation of Eq. (\ref{eq:Conductivity equation}),
a semi-linear model that is obtained after simplification and taking
log of both sides of Eq. (\ref{eq:Conductivity equation}) one obtains

\begin{equation}
h^{*}(t)=B^{*}+f(D,t)\:,\label{eq:Conductivity equation-reforming-1}
\end{equation}
where 

\[
h^{*}(t)=\log\left\{ h(t)\right\} ,\ B^{*}=\log(B),
\]
and

\[
f(D,t)=\log\left(1-\sum_{m=1}^{M}\frac{8}{\pi^{2}(2m-1)^{2}}\exp\{-(2m-1)^{2}\pi^{2}D\frac{t}{L^{2}}\}\right)
\]
which makes $h^{*}(t)$, a linear function of $B^{*}$ while $f(D,t)$
is known to be rather small. Note that $\log(1-e^{-x})\rightarrow e^{-x}$
as $x\rightarrow\infty$ and therefore $f(D,t)$ decays exponentially
for large with increasing time which makes the linearization with
respect to $D$ undoable.

\section{\label{sec:Results-and-Discussions}Results and Discussions}

The accuracy of the series in Eq. (\ref{eq:Conductivity equation})
was studied in this work to reduce the computational time, it showed
that the number of terms $m$ may be chosen to be around 200 terms
to obtain an accuracy of 99.9\% in the middle of the sample at $t=0$.
Already at the second measurement at $t=0.003L^{2}/D$ only four terms
are required to obtain the same accuracy. For systems with large amounts
of data, a strategy for the selection of the number of terms in the
series could save considerable computation time. However, here this
is not the primary focus and the the calculations where not very much
time consuming, which is why all calculations where made using $M=200$
terms.

Measurements from known $D$ and $B$ using $w_{D}=0.1D$, $w_{B}=0.1B$,
and $v=0.1z_{av}$ where $z_{av}$ is the average of all generated
measurements \textcolor{black}{i.e $z_{av}=\frac{1}{N}\sum z_{i}$}.
Then the Kalman filer was used to obtain approximations of $D$ and
$B$. This was done first for the non-linear model Eq. (\ref{eq:Conductivity equation}),
and then for the partly linearized model Eq. (\ref{eq:Conductivity equation-reforming-1}).
To compare the accuracy in finding the least square error of the predicted
measurements based on different state parameters found using model
Eq. (\ref{eq:Conductivity equation}) and model Eq. (\ref{eq:Conductivity equation-reforming-1})
respectively, the least square error $\mathcal{M}(x)$ was calculated
using Eq. (\ref{eq:def_square_error}). The non-linear model gave
$\mathcal{M}(x)=15\%$ while the partly linearized model gave 30\%.
For the reason of this, the partly linearized model is given up and
the model Eq. (\ref{eq:Conductivity equation}) is used in the continued
analysis. 

\begin{figure}[H]
\begin{centering}
\includegraphics[width=0.6\textwidth,height=0.6\textwidth]{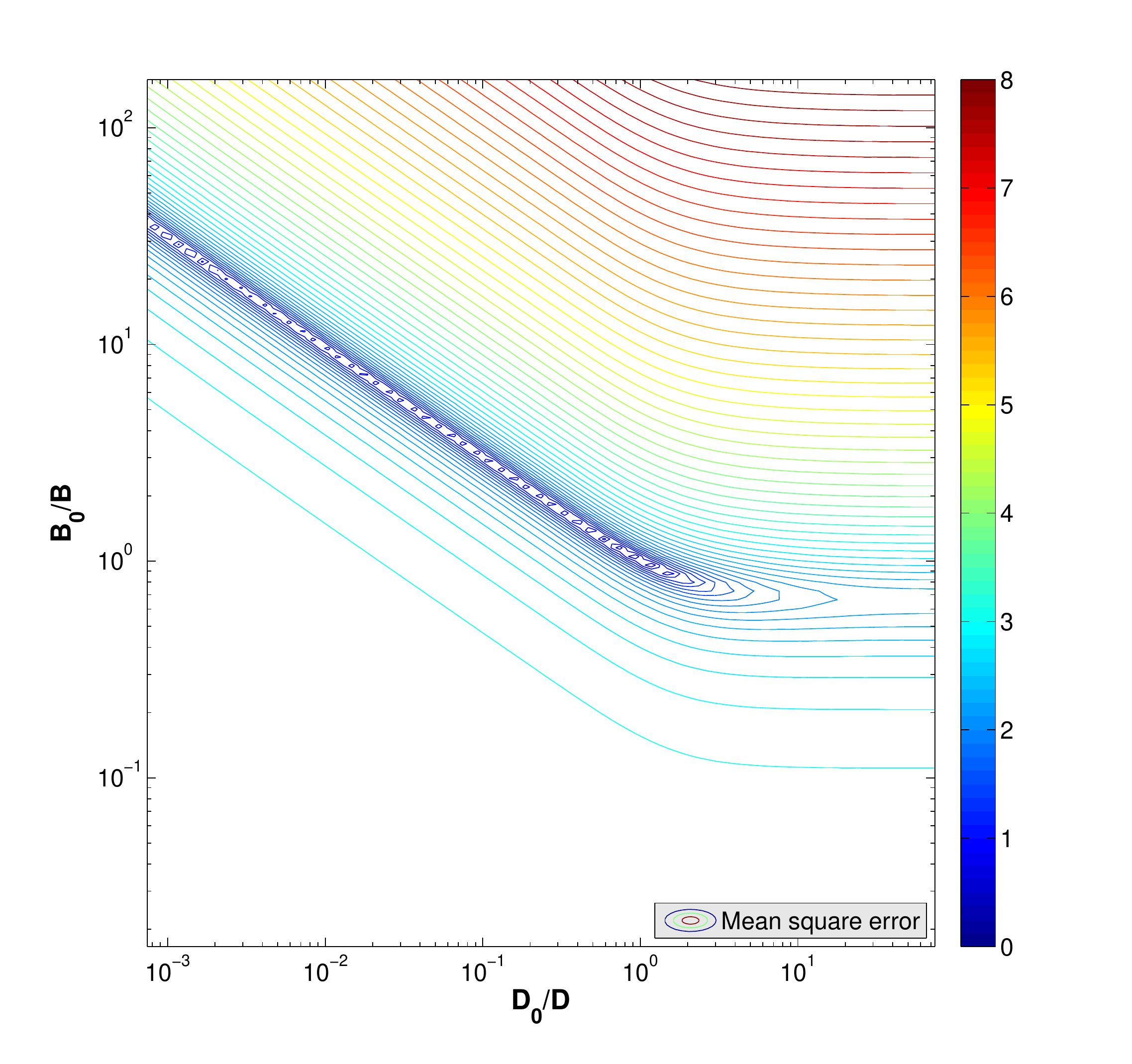}
\par\end{centering}

\protect\caption{The mean square error of the conductivity predictions for different
initial values of the relative initial parameter values $D_{0}/D$
and $B_{0}/B$. The noises are in the range \textpm 5\% for parameters
and \textpm 10\% for measurements. }
\label{fig:figure 1}
\end{figure}

Assuming that the information of the parameters $D$ and $B$ is uncertain,
the Kalman filtering has to converge from initial values that are
several orders of magnitudes different from the true values. To explore
what this means, the mean square error is calculated for a very wide
range of initial values. Mean square error here refers to the error
of calculated estimates of the initial guesses $D_{0}$ and $B_{0}$
directly compared with the true values of $D$ and $B$. No iterations
are made. To cover a large variety of initial values a mesh 101 values
for $D$ and for each 101 values for $B$ are used. The mesh covers
values of $D_{0}$ ranging from $0.001D$ to $100D$ and values of
$B_{0}$ from $0.01B$ to $100B$. \autoref{fig:figure 1} shows the
mean square error for the initial values of $D_{0}$ and $B_{0}$
for the constructed mesh. 

Regions with small gradients are found along a line for which $D_{0}/D$
approximately equals $(B_{0}/B)^{-2}$. The several local minima along
this trajectory correspond to the resolution of the grid and are merely
graphical anomalities. Also in the region of small $D_{0}$ and small
$B_{0}$ the gradient is very small. This makes the convergence rate
of any gradient driven algorithm small. However, no local minimum
seems to be present and convergence should be possible in entire range
of initial values even if the convergence may be very slow in the
above describe regions. A clear minimum mean square error is found
around the close to $D$ and $B$. The result is strongly influenced
by the rapid changes due to the exponential behavior of Eq. (\ref{eq:Conductivity equation}). 

For the study of the Kalman filter measurements $z$ are generated
for 250 measurements in the time interval $0\leq t\leq11.7L^{2}/D$.
In this study, $D$ and $B$ are known parameters, but in a real case
they are not known but believed to be in the neighborhood of the \textit{a
priori} guess. As it is suggested by \autoref{fig:figure 1}, a very
large variation of convergence rates are anticipated. 

The Kalman filter is studied by gathering 10 generated experiments
together, each experiment has 25 measurements, and each measurement
has random noises of \textpm 5\% for parameters and \textpm 10\%,
\textpm 50\%, and \textpm 100\% for measurements, which constructs
a measurement vector $z_{i}$, i=1,2.3.....,250. The initial selected
state variables are chosen from $0.1D$ to $10D$ for $D_{0}$ and
from $0.1B$ to $10B$ for $B_{0}$.

The initial parameters error $\boldsymbol{P}_{0}^{-}$ is selected
according to Eq. (\ref{eq:pred-param-error}) as follows

\begin{gather}
\boldsymbol{P}_{0}^{-}=\left[\begin{array}{cc}
9.9^{2}D_{0}^{2} & 0\\
0 & 9.9^{2}B_{0}^{2}
\end{array}\right],\label{eq:Initial covaraince error}
\end{gather}
The $\boldsymbol{P}_{k}^{-}$ is updated using Eq. (\ref{eq:P-recursion-2}).

\subsection{Different {\boldmath{$R$}} and {\boldmath{$Q$}}$_{k}$}

First $\boldsymbol{R}^{(i)}$ values are calculated as in Eq. (\ref{eq:noise max-1}).\textcolor{black}{{}
So, }$\boldsymbol{R}^{(i)}$\textcolor{black}{{} is the squared largest
element of the variance encountered so far, i.e. so far means before
iteration $k$.}

The effect of $\boldsymbol{Q}_{k}$ is studied by using, 

\begin{equation}
\boldsymbol{Q}_{k}=\left[\begin{array}{cc}
(D_{k}-D_{0})^{2} & 0\\
0 & (B_{k}-B_{0})^{2}
\end{array}\right]\,.\label{eq:covariance parameters error D B}
\end{equation}
This means that $\boldsymbol{Q}_{k}$ increases as the iterations
proceed. The expected effect is that the convergence rate increases
with increasing $\boldsymbol{Q}_{k}$. Which is selected to speed
up the convergence rate.

\autoref{fig: figure2}(a) shows a color plot of the obtained values
for $D_{1}/D$, here 1 is number of iterations. The markers ($\boldsymbol{\times}$)
that are included show the obtained values of $D_{1}/D$ and $B_{1}/B$
using Kalman filter, for 41$\times$41 initial values $D_{0}/D$ and
$B_{0}/B$. The obtained\textit{ $(D_{1}/D$,$B_{1}/B)$} outside
the selected range are excluded from convergence plot to make it in
the same range as the color plot. The white areas in the figure give
21.5\% of the obtained $D_{1}/D$ that found to have negative values
with no physical meaning. Consequently, this leads to divergent result
in following iteration since the exponential term in the diffusion
model would have large positive values. A large step for Kalman gain
is $\mathbf{K}{}_{k}$ seems to be the reason, which can be forced
to be small by assuming $\boldsymbol{R}$ large, but the rate of convergence
would be slow. 

\begin{center}
\begin{figure}[H]
\begin{centering}
\includegraphics[width=0.5\textwidth,height=0.5\textwidth]{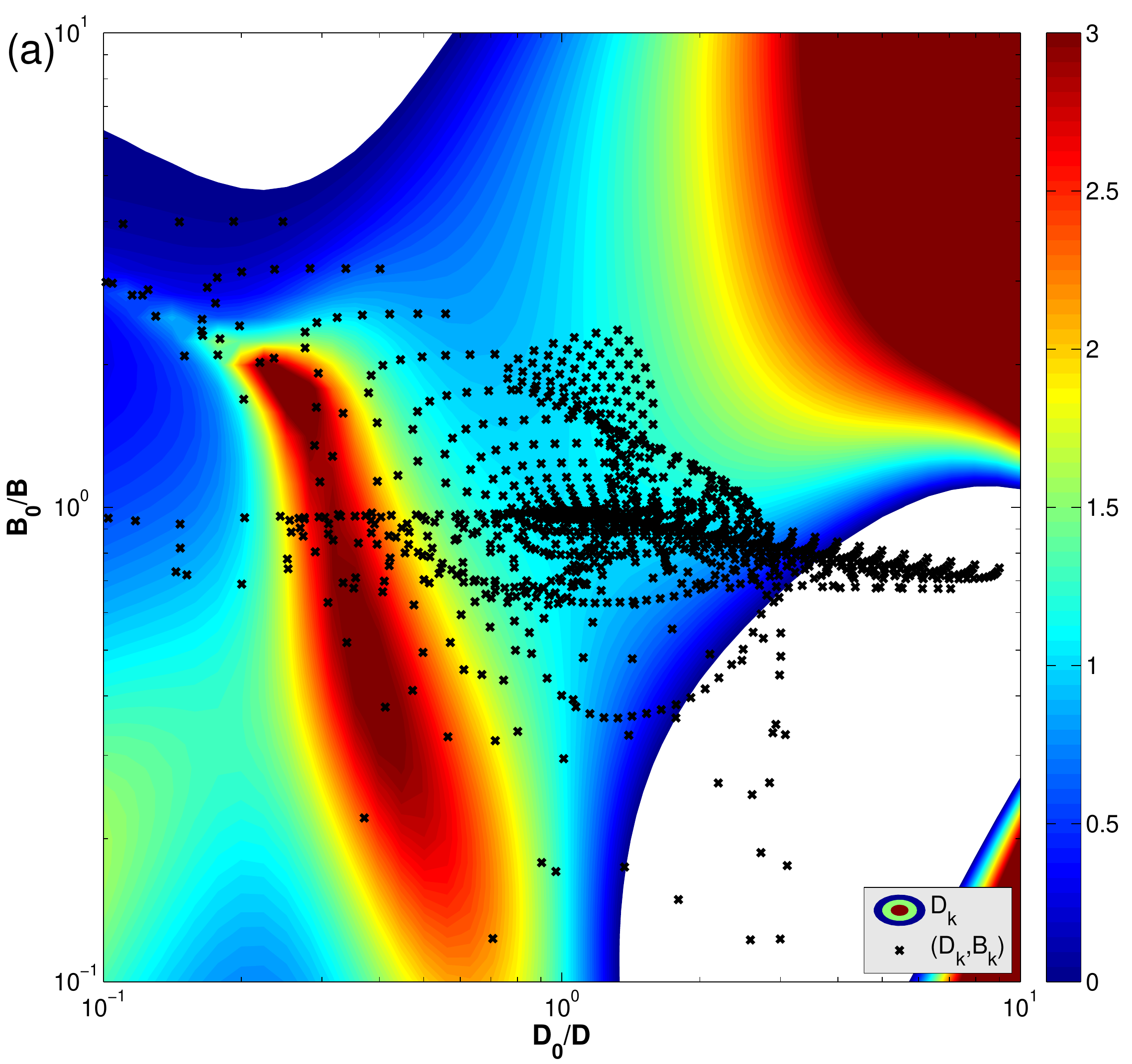}\includegraphics[width=0.5\textwidth,height=0.5\textwidth]{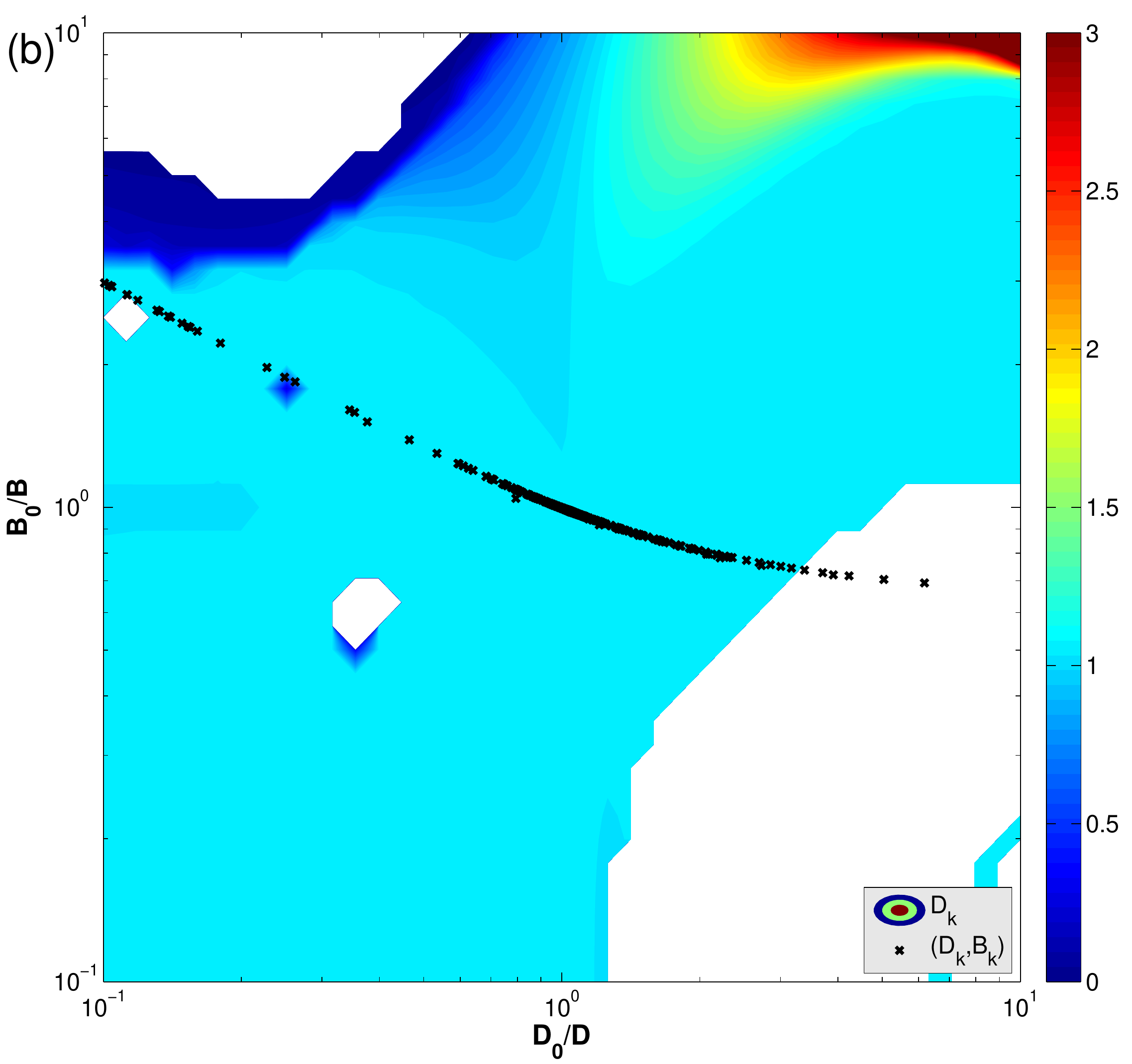}
\par\end{centering}

\protect\caption{Color plot of $D_{k}$ as a function of $D_{0}$ and $B_{0}$. The
markers ($\times$) show the position of the resulting $D_{k}$ and
$B_{k}$ for the 1681 initial starting points in the range. (a) $k=1$
iteration, and (b) $k=20$ iterations. Noises are \textpm 5\% for
the parameters and \textpm 10\% for the measurements.}
\label{fig: figure2}
\end{figure}

\par\end{center}

\autoref{fig: figure2}(b) shows the color and convergence plot after
20 iterations. It can be seen that using larger values for $\boldsymbol{Q}_{k}$
increases the rate of convergence but the divergent cases increased
from 21.5\% to 22.4\%. 

\begin{figure}[H]
\begin{centering}
\includegraphics[width=0.5\textwidth,height=0.5\textwidth]{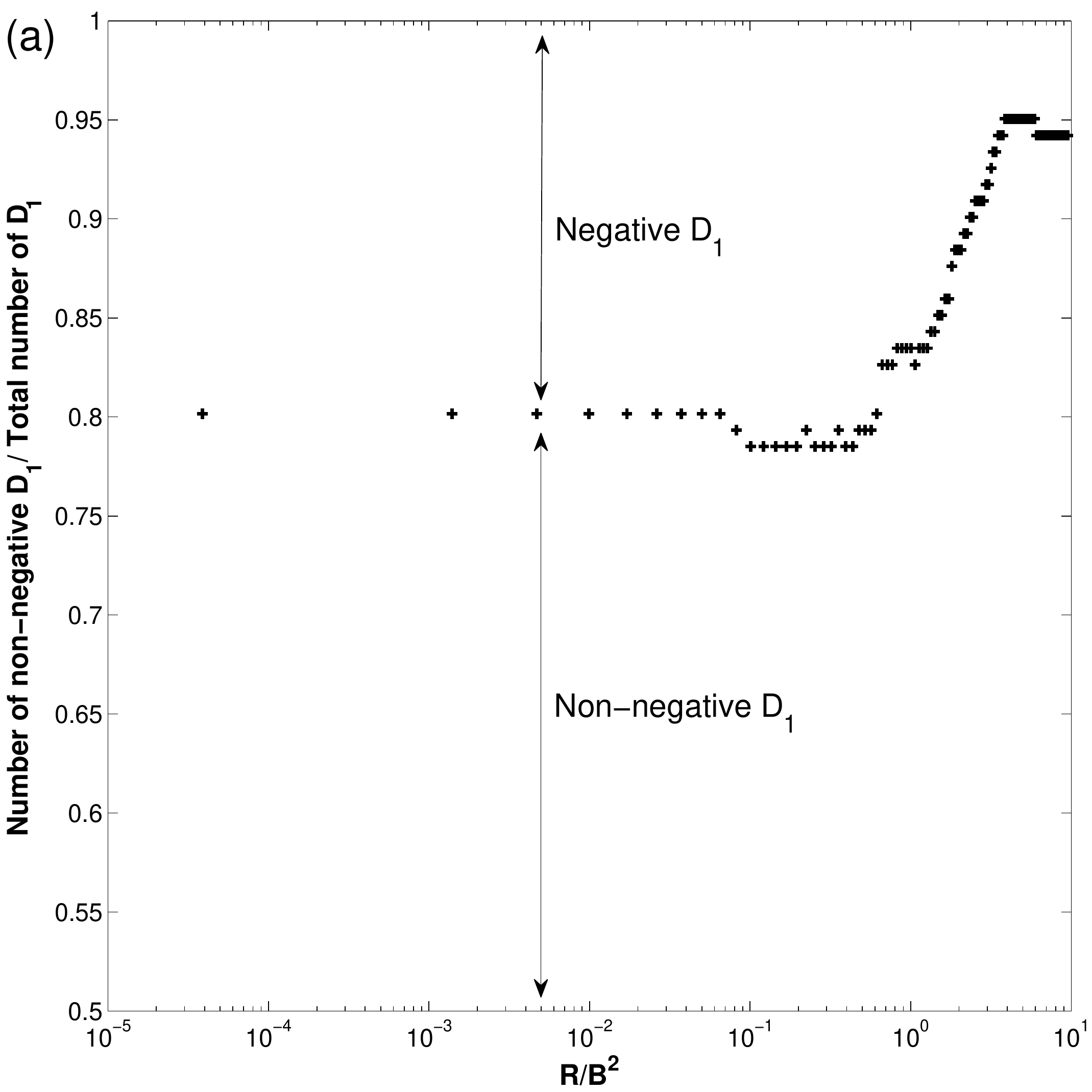}\includegraphics[width=0.5\textwidth,height=0.5\textwidth]{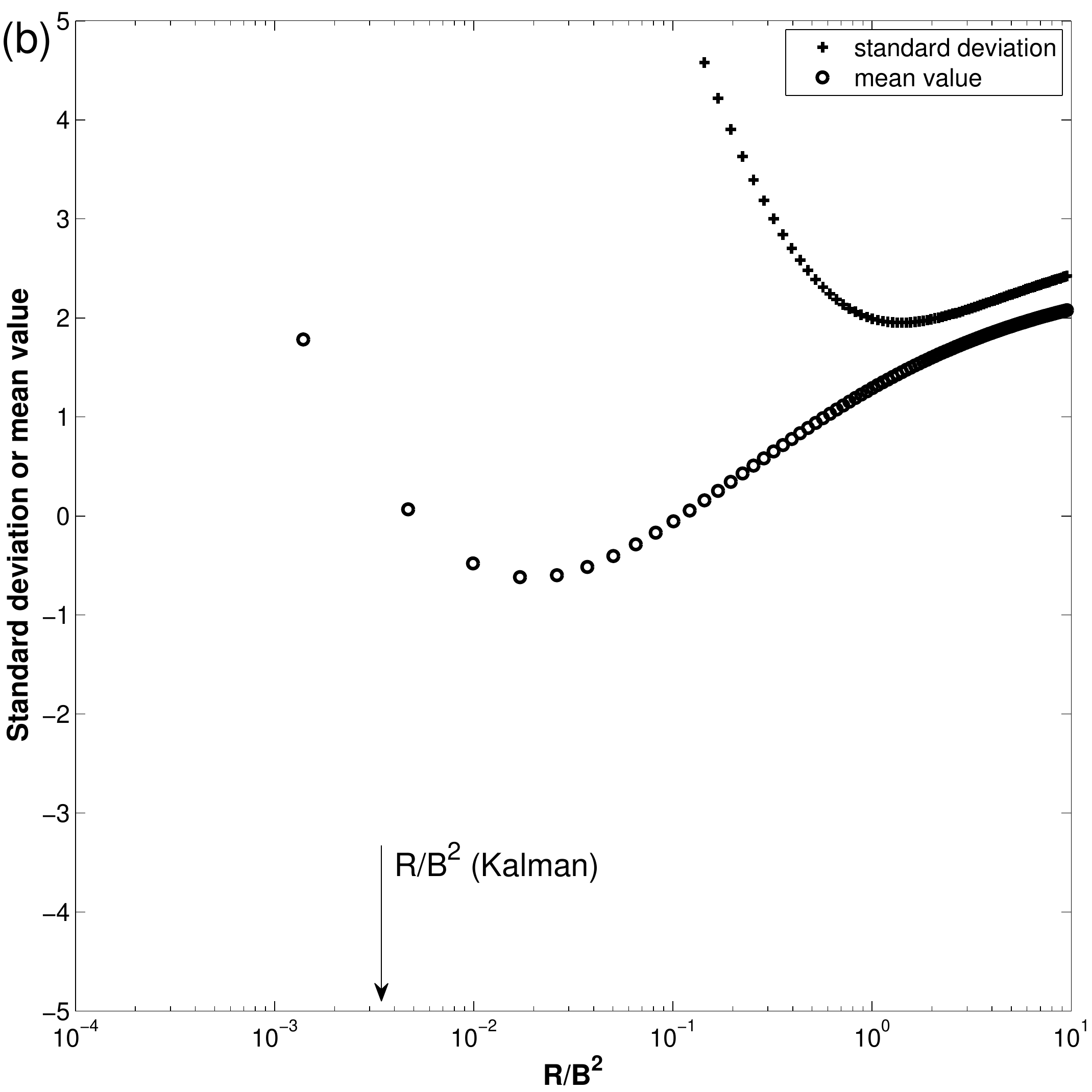}
\par\end{centering}

\protect\caption{Effect of $\boldsymbol{R}$ on $D_{1}/D$. (a) convergent of $D_{1}/D$
over total number of $D_{1}/D$. (b) standard deviation and mean value,
the standard deviation values over 5 are excluded from the graph.
$\boldsymbol{Q}=0$.}
\label{fig:figure3}
\end{figure}

The effect of $\boldsymbol{R}$ values on the convergent $D_{1}/D$
is shown in \autoref{fig:figure3}(a). The figure shows that the percentage
of the number of convergent $D_{1}/D$ over the total number of $D_{1}/D$
almost around 80\% as the $\boldsymbol{R}$ value increases from 0.000039$B^{2}$
to 0.6$B^{2}$, after that it begins to increase to 95\% as the $\boldsymbol{R}$
increases to 3$B^{2}$, then it almost keep around this percentage
with larger values for $\boldsymbol{R}$. A possible reason is that
the Kalman gain step becomes small as the $\boldsymbol{R}$ value
increases.

The effect of $\boldsymbol{R}$ on standard deviations and mean values
are shown in \autoref{fig:figure3}(b). The figure shows that the
standard deviation decreases as the $\boldsymbol{R}$ increases untill
$\boldsymbol{R}=1.5B^{2}$, then it increases slowly, also the mean
value decreases as the $\boldsymbol{R}$ increases to 0.02$B^{2}$,
and then increases again. The $\boldsymbol{R}/B^{2}$ (Kalman) in
the graph denoted suggested $\boldsymbol{R}$ for ordinary Kalman
filter based on the average value of $z$ and the 10\% measurements
noise, i.e $\boldsymbol{R}=(0.1z_{av})^{2}$, which is not the optimal
case. One reason is that standard deviations and mean values are computed
after one iteration, also the negative values of $D$ are included.
Slightly different results is obtained after more iterations, reason
should be the non-linear model.

\begin{figure}[H]
\begin{centering}
\includegraphics[width=0.5\textwidth,height=0.5\textwidth]{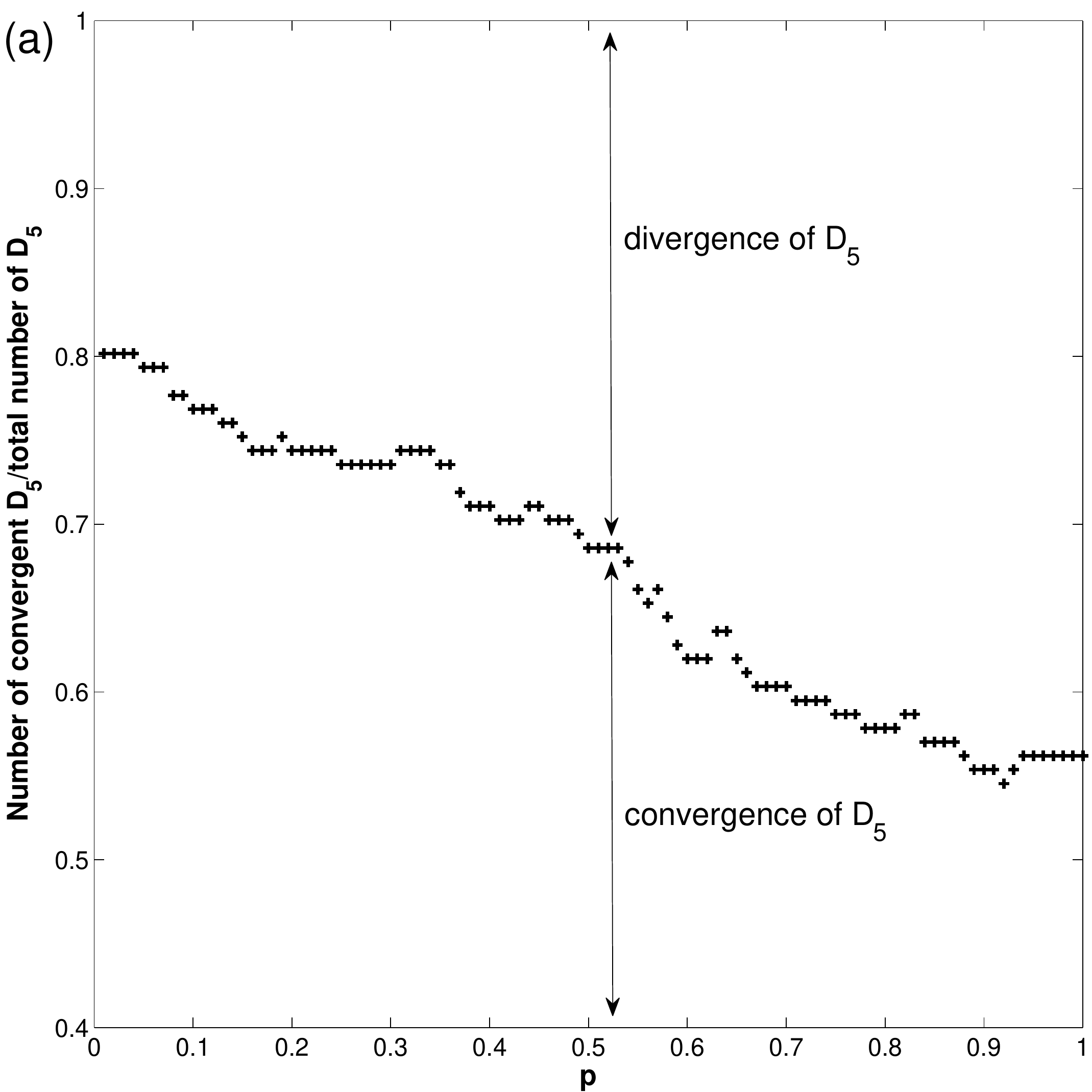}\includegraphics[width=0.5\textwidth,height=0.5\textwidth]{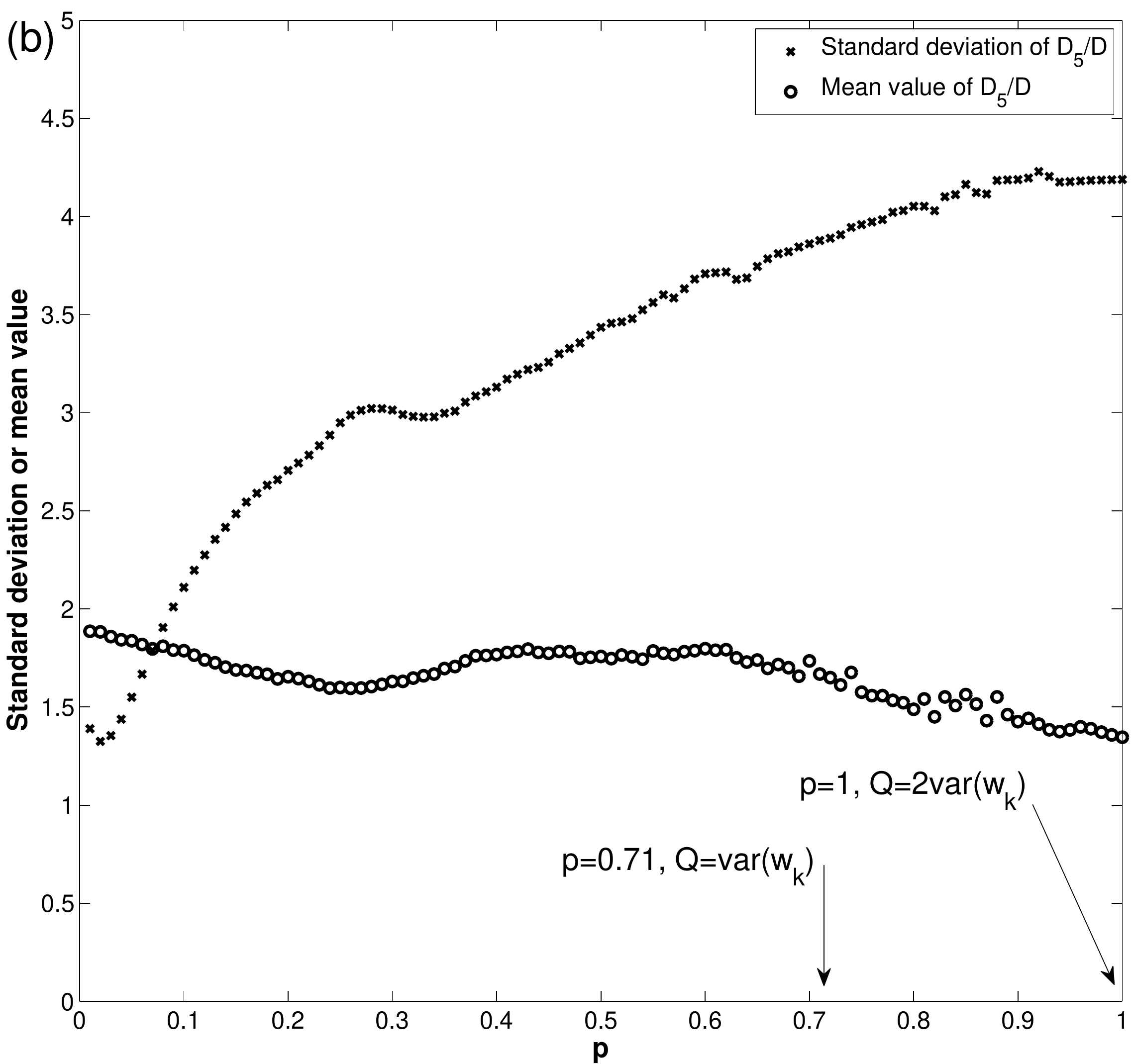}
\par\end{centering}

\protect\caption{The effect of $p$ on $D_{5}/D$. (a) on convergent of $D_{5}/D$
over total number of $D_{5}/D$ (b) standard deviation and mean value
of $D_{5}/D$. $\boldsymbol{R}$ is chosen as in Eq. (\ref{eq:noise measur-1}). }
\label{fig:figure4}
\end{figure}

\autoref{fig:figure4}(a) shows the effect of different values of
$\boldsymbol{Q}$ using $0.01\leq p\leq1$ and $\boldsymbol{R}$ as
given in Eq. (\ref{eq:noise measur-1}) on the convergent numbers
of $D_{5}/D$ over total number of $D_{5}/D$. It is clearly seen
that the number of convergent $D_{5}/D$ decreases as the $p$ increases.
Using large $\boldsymbol{Q}$ with small $\boldsymbol{R}$ affects
the Kalman gain stability, this could be a possible reason.

\autoref{fig:figure4}(b) shows the effect of $p$ on the standard
deviations and mean values for the obtained $D_{5}/D$. The figure
shows that standard deviation increases as $p$ increases, and mean
values play around 1.7 as the $p$ increases. One reason could be
that using large values of $\boldsymbol{Q}$ with small values for
$\boldsymbol{R}$ increases the Kalman gain step. The $\boldsymbol{Q}$
value at $p=0.71$, random walk (cf. \citep{Brown(1983)}), and at
$p=1$ , the suggested variance as in Eq. (\ref{eq:updated value of Q}),
does not give significant effect on both standard deviation and mean
value. A non-linear model might be a reason, so it is expected to
have effect with more iterations. The $p$ value is chosen for 0.01
that give smallest standard deviation and largest percentage of number
of convergent $D$ over total number of $D$ to be one method (see
\autoref{Table:1}).

\subsection{Large {\boldmath{$R$}}  and {\boldmath{$Q$}} }

\begin{figure}[H]
\begin{centering}
\includegraphics[width=0.5\textwidth,height=0.5\textwidth]{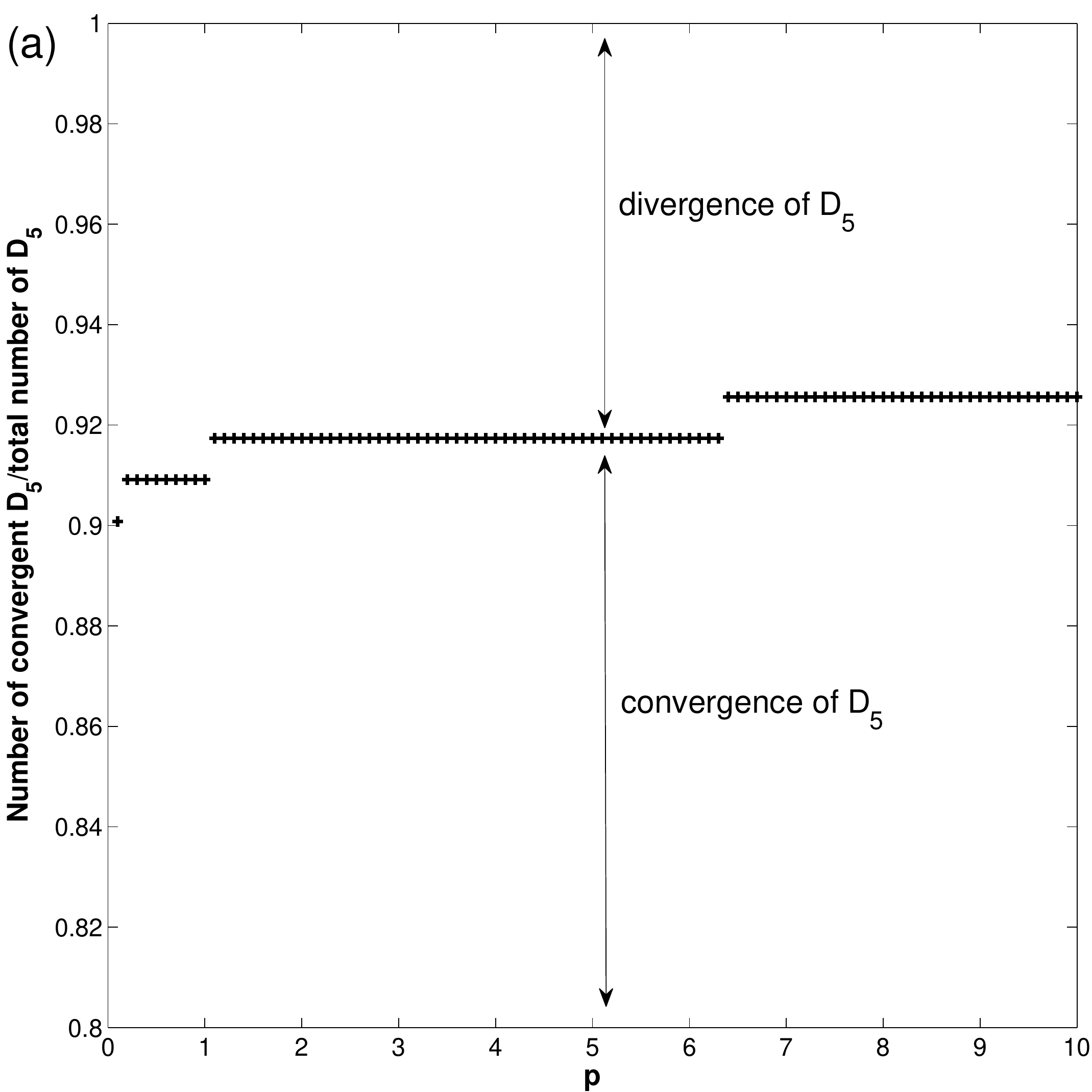}\includegraphics[width=0.5\textwidth,height=0.5\textwidth]{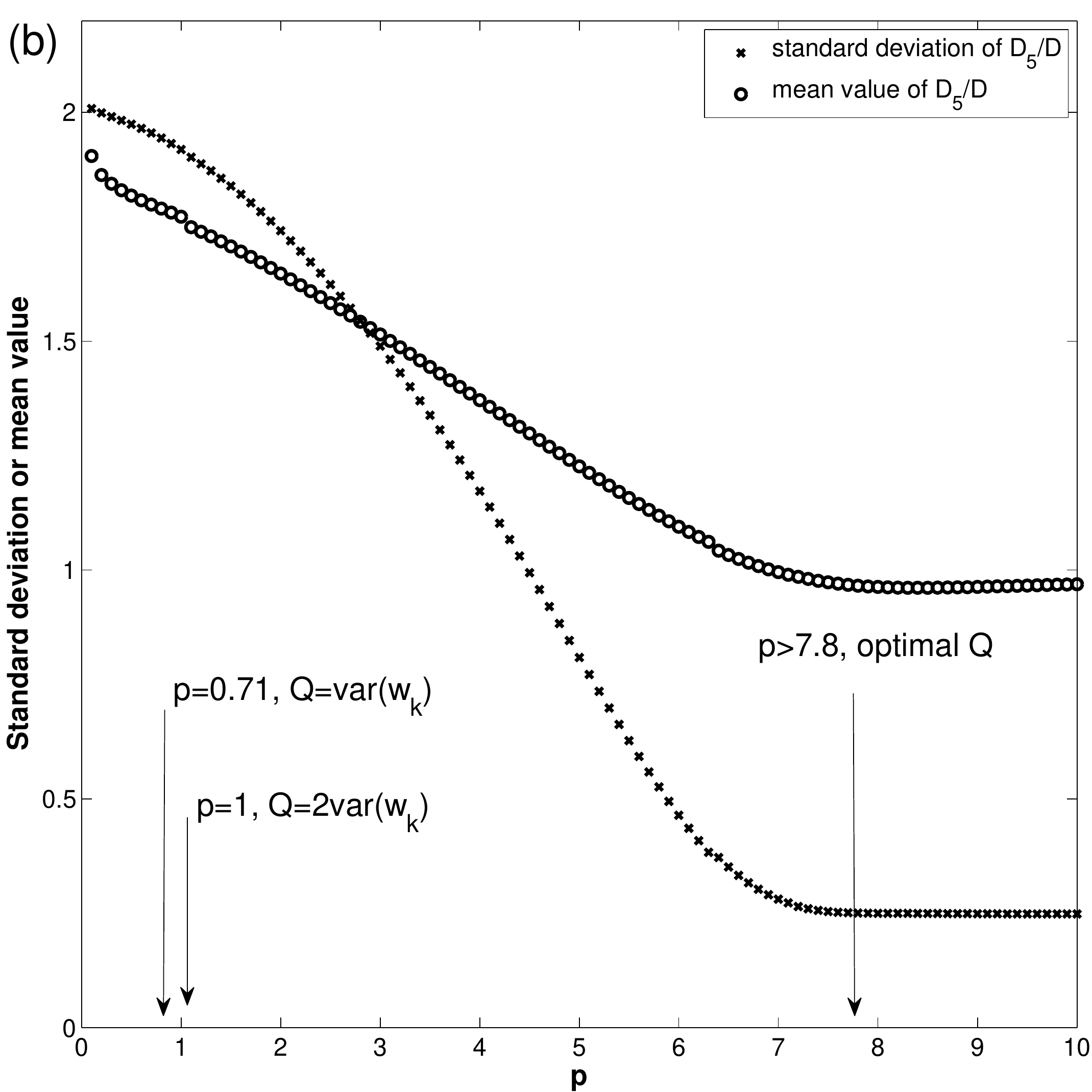}
\par\end{centering}

\protect\caption{The effect of $p$ on $D_{5}/D$ using $\boldsymbol{R}$ large. (a)
convergent of $D_{5}/D$ over total number of $D_{5}/D$ (b) standard
deviation and mean value of $D_{5}/D$. The large $\boldsymbol{R}$
is chosen as in Eq. (\ref{eq:covariance measurements error D B RLQL}).}
\label{fig: figure5}
\end{figure}

By studying Eqs (\ref{eq:Kalman gain-2}) and (\ref{eq:P-recursion-2}),
and \autoref{fig:figure3} it is believed that choosing $\boldsymbol{R}$
large, decreases the risk of ending up with divergent results for
$D_{k}$. \autoref{fig: figure5}(a) shows that increasing $p$ for
large values using $\boldsymbol{R}$ large increases the percentage
of the number of convergent $D_{5}$ by 3\%, this is an indication
that using $p>0$ with $\boldsymbol{R}$ large covers a wide range
of initial parameters that converges. 

The effect of $\boldsymbol{Q}$ on the standard deviations and mean
values of the obtained $D_{5}/D$ are shown in \autoref{fig: figure5}(b).
It can be seen that the standard deviation and mean value decrease
as the $p$ increases, untill optimal is obtained with $p>7.8$. Also,
increasing $p$ speeds the rate of convergence. Again, this could
be the effect of non-linearity of the model. 

It can be concluded that using $\boldsymbol{R}$ and $\boldsymbol{Q}$
large increase the possibility of convergence for a wide range of
$D_{0}$ and $B_{0}$ and speeds the convergence of parameters. To
choose $\boldsymbol{R}$ for large value, a vector $v_{s}$ contains
the largest variance value between measurement, $z,$ and predicted
, $h(x_{0})$, for each initial parameters is chosen as

\begin{equation}
v_{s}=[v^{(1)}v^{(2)}....v^{(1681)}]^{T}.\label{eq:noise 1681}
\end{equation}
where $s$ is the number of initial parameters, i.e $s=1,2,....,1681$,
and $v^{(i)}$ values are given by Eq. (\ref{eq:noise max}), then
the $\boldsymbol{R}$ value is chosen as the maximum value in the
whole initial parameters combinations as

\begin{gather}
\boldsymbol{R}=\max(v_{s})I,\label{eq:covariance measurements error D B RLQL}
\end{gather}

To choose $\boldsymbol{Q}$ large, the $\boldsymbol{Q}$ is chosen
to be equal to $\boldsymbol{P}_{0}^{-}$ as in Eq. (\ref{eq:Initial covaraince error}),
since the $\boldsymbol{Q}$ represents the variance between the seeking
parameters and instant parameters. 

\begin{figure}[H]
\begin{centering}
\includegraphics[width=0.5\textwidth,height=0.5\textwidth]{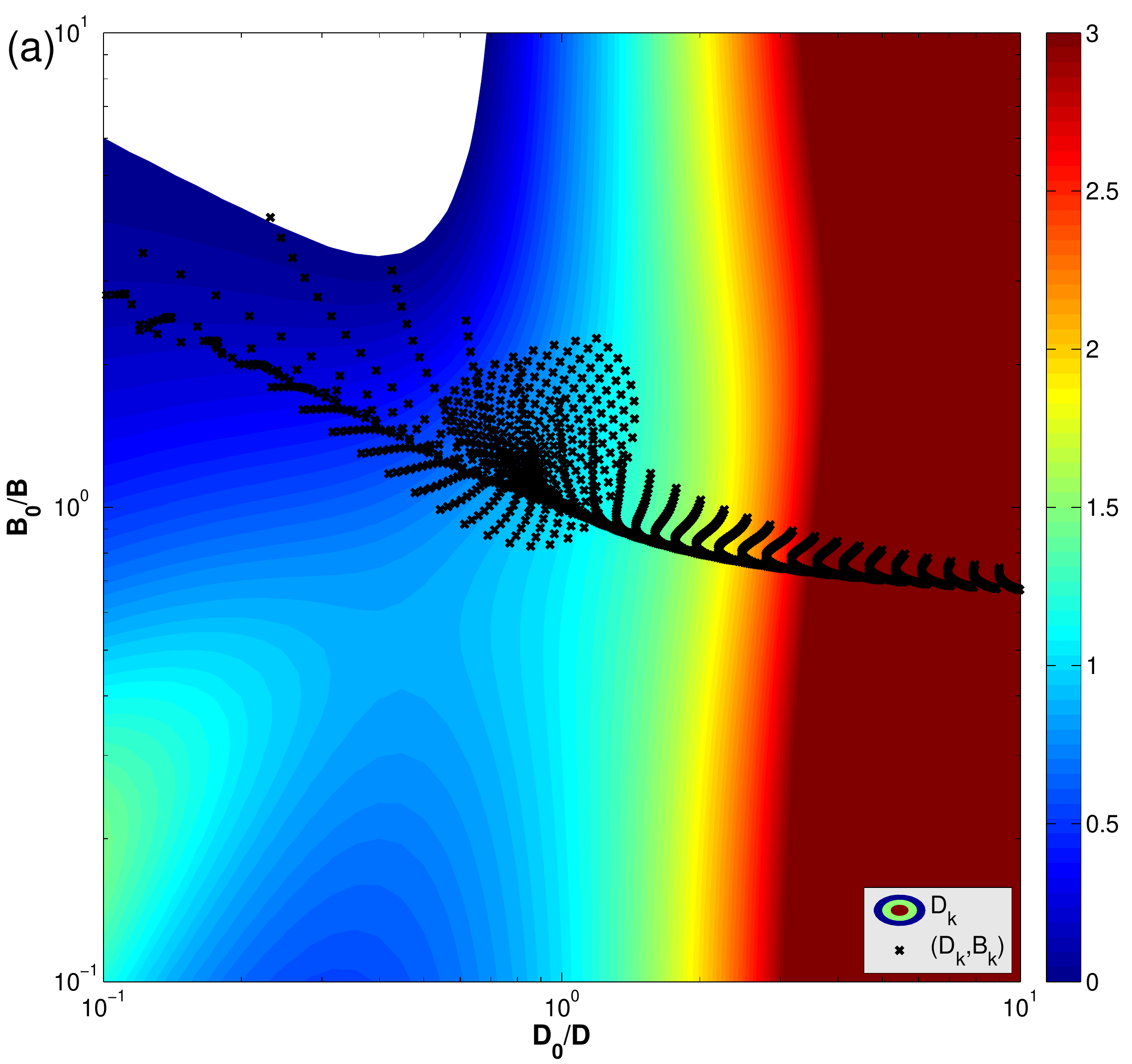}\includegraphics[width=0.5\textwidth,height=0.5\textwidth]{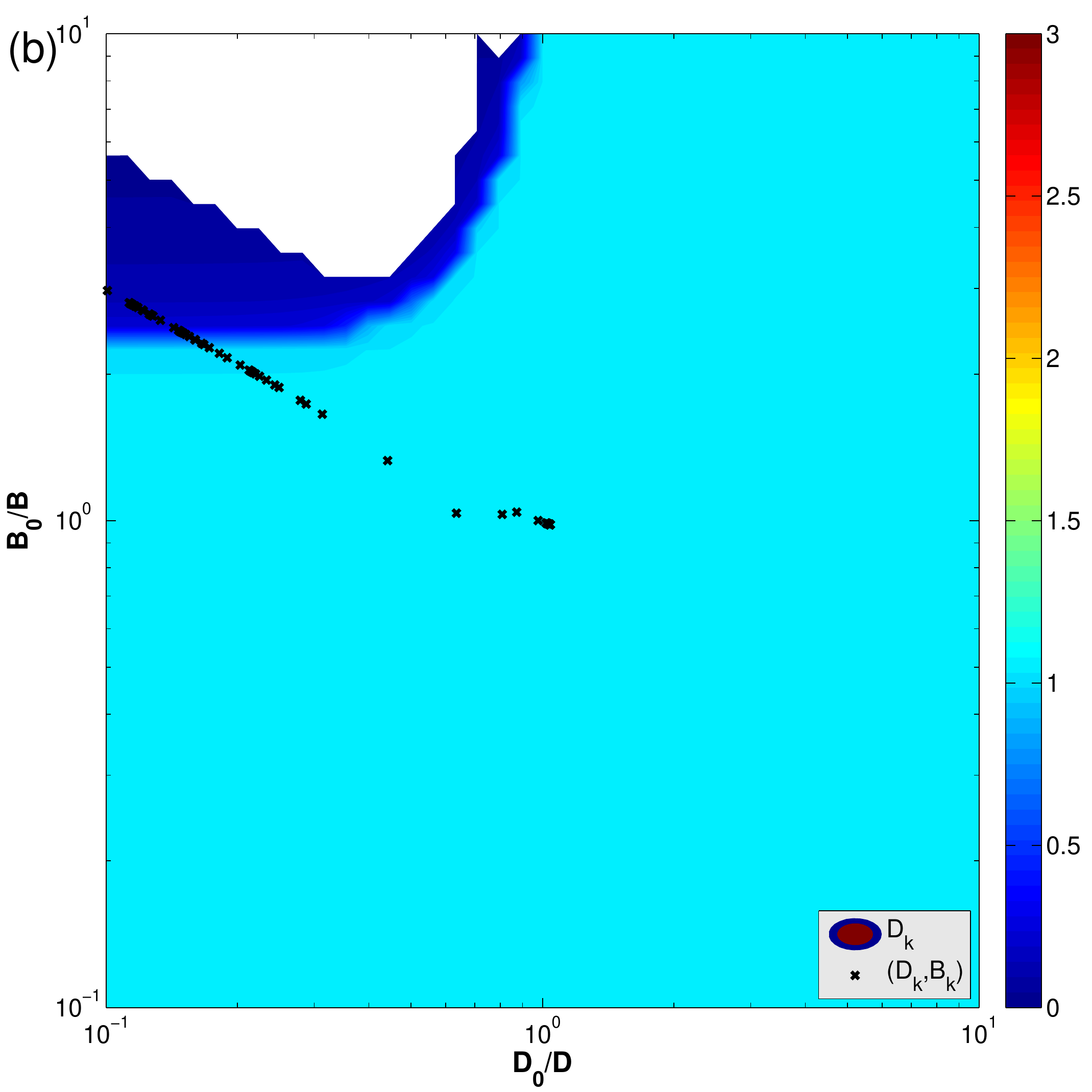}
\par\end{centering}

\protect\caption{Color plot for $D_{k}$ merged with convergence plot between $(D_{k}$,$B_{k})$
for initial starting points $D_{0}$and $B_{0}$ after (a) 1 iteration,
for large $\boldsymbol{R}$ and $\boldsymbol{Q}=0$, and (b) 20 iterations
for large $\boldsymbol{R}$ and large $\boldsymbol{Q}$. $\boldsymbol{Q}$
is selected to be large and equal to $\boldsymbol{P}_{0}^{-}$ as
in Eq. (\ref{eq:Initial covaraince error}), while $\boldsymbol{R}$
is chosen as in Eq. (\ref{eq:covariance measurements error D B RLQL}).
Noises are \textpm 5\% for the parameters and \textpm 10\% for the
measurements. }
\label{fig: figure6}
\end{figure}

\autoref{fig: figure6}(a) shows that the divergent result after 1
iteration using large $\boldsymbol{R}$ and $\boldsymbol{Q}=0$ decreased
to 7.8\% compared with the divergence result of 21.5\% for different
$\boldsymbol{R}$ values as shown in \autoref{fig: figure2}(a). It
also decreased to 7.9\% after 20 iterations using large $\boldsymbol{R}$
and $\boldsymbol{Q}$ as shown in \autoref{fig: figure6}(b) compared
to the divergence result of 22.4\% for different $\boldsymbol{R}$
and $\boldsymbol{Q}$ values as shown in \autoref{fig: figure2}(b). 

It can be discussed that large values of $\boldsymbol{R}$ and $\boldsymbol{Q}$
decreases the percentage of divergent result of $D_{k}$, and also
speeds the rate of convergence as shown in \autoref{fig: figure2}
and \autoref{fig: figure6}.

\subsection{Different methods }

\begin{figure}[H]
\begin{centering}
\includegraphics[width=0.5\textwidth,height=0.5\textwidth]{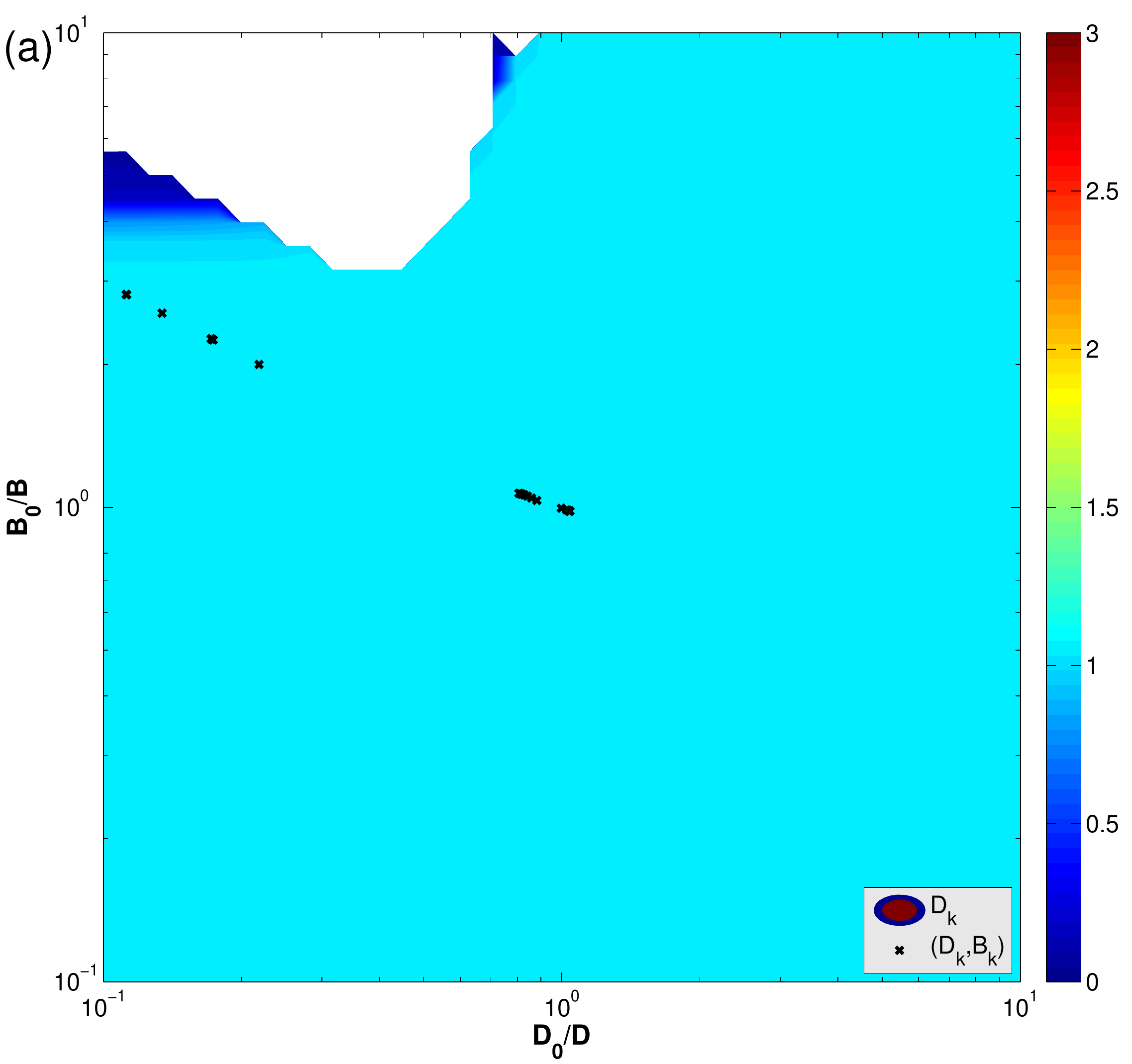}\includegraphics[width=0.5\textwidth,height=0.5\textwidth]{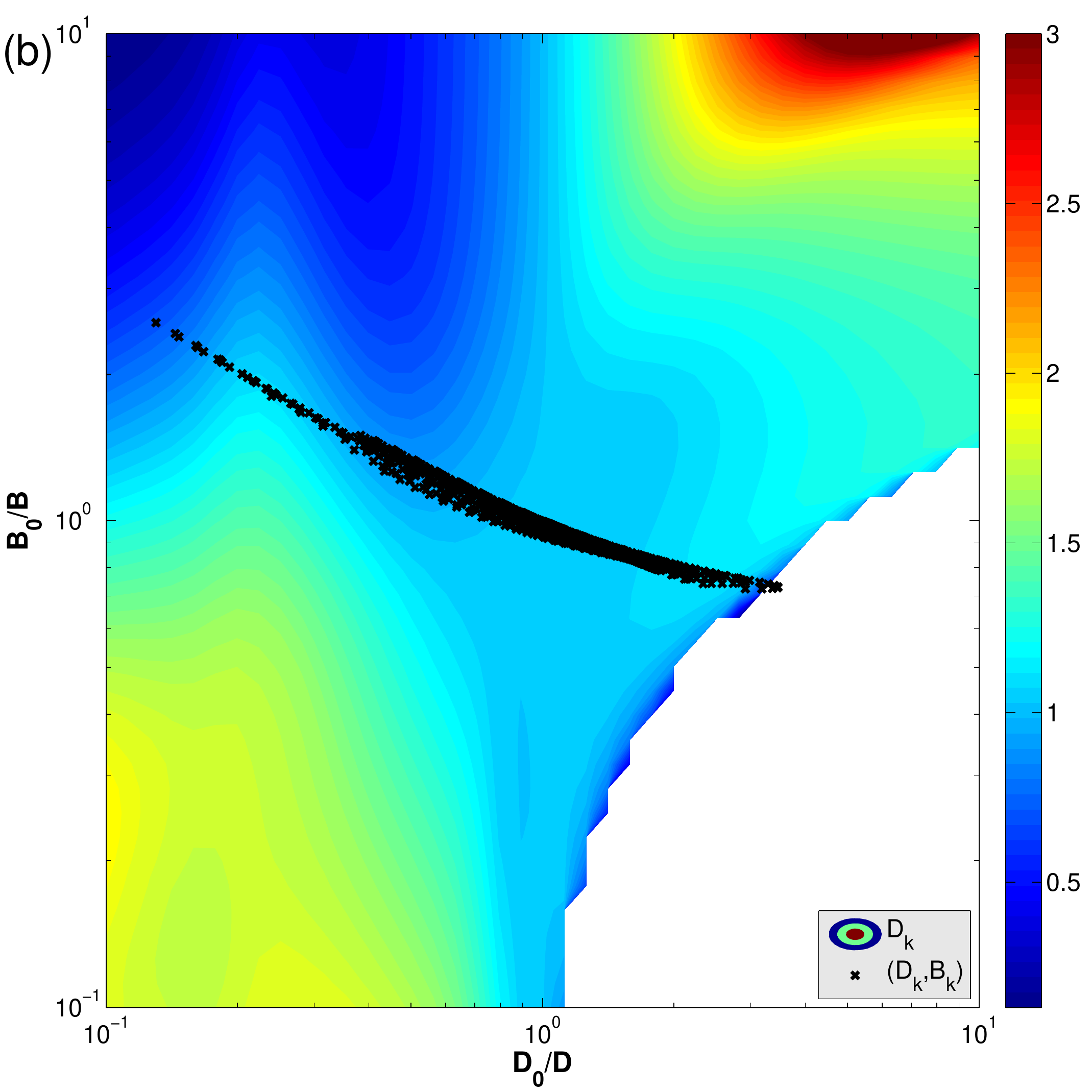}
\par\end{centering}

\begin{centering}
\includegraphics[width=0.5\textwidth,height=0.5\textwidth]{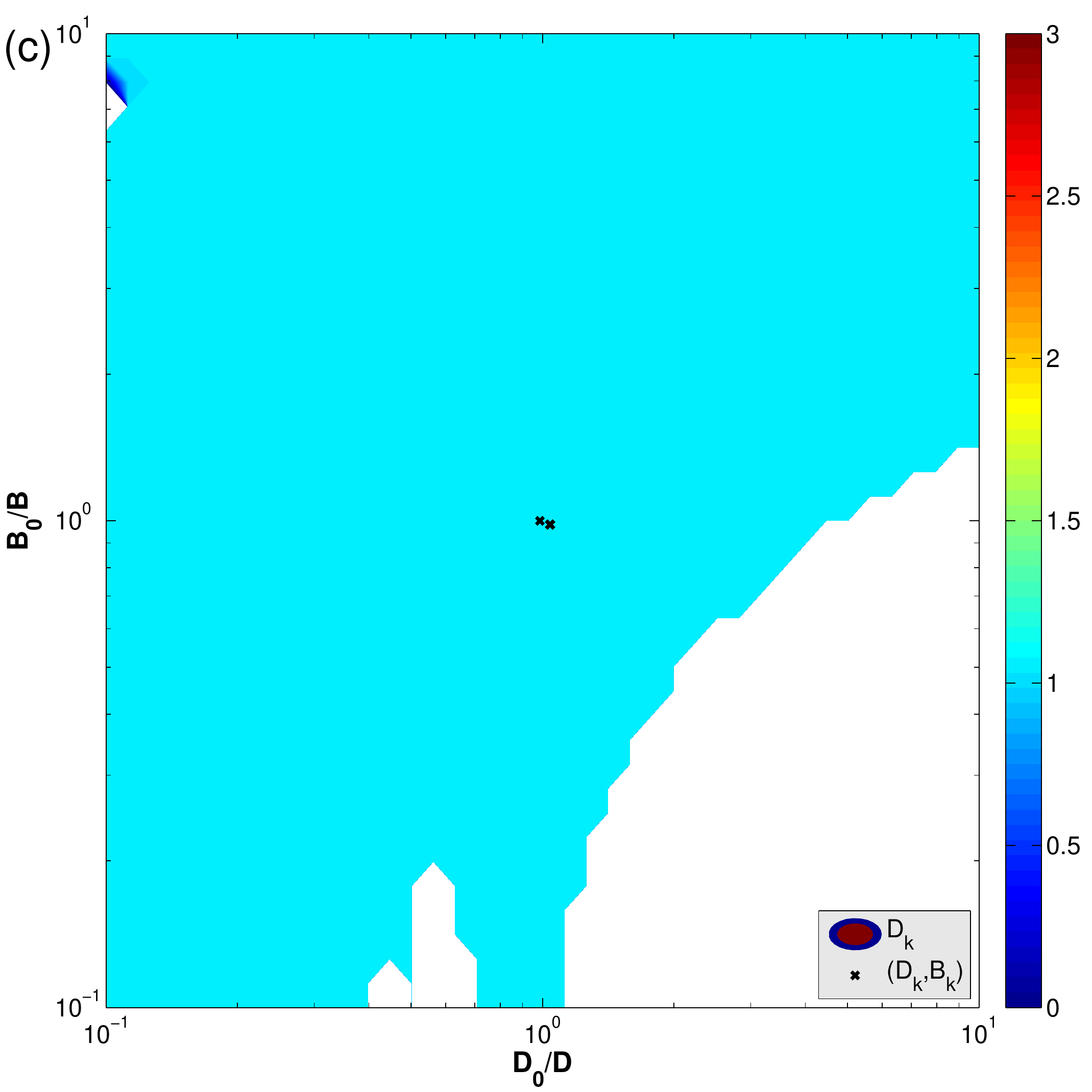}\includegraphics[width=0.5\textwidth,height=0.5\textwidth]{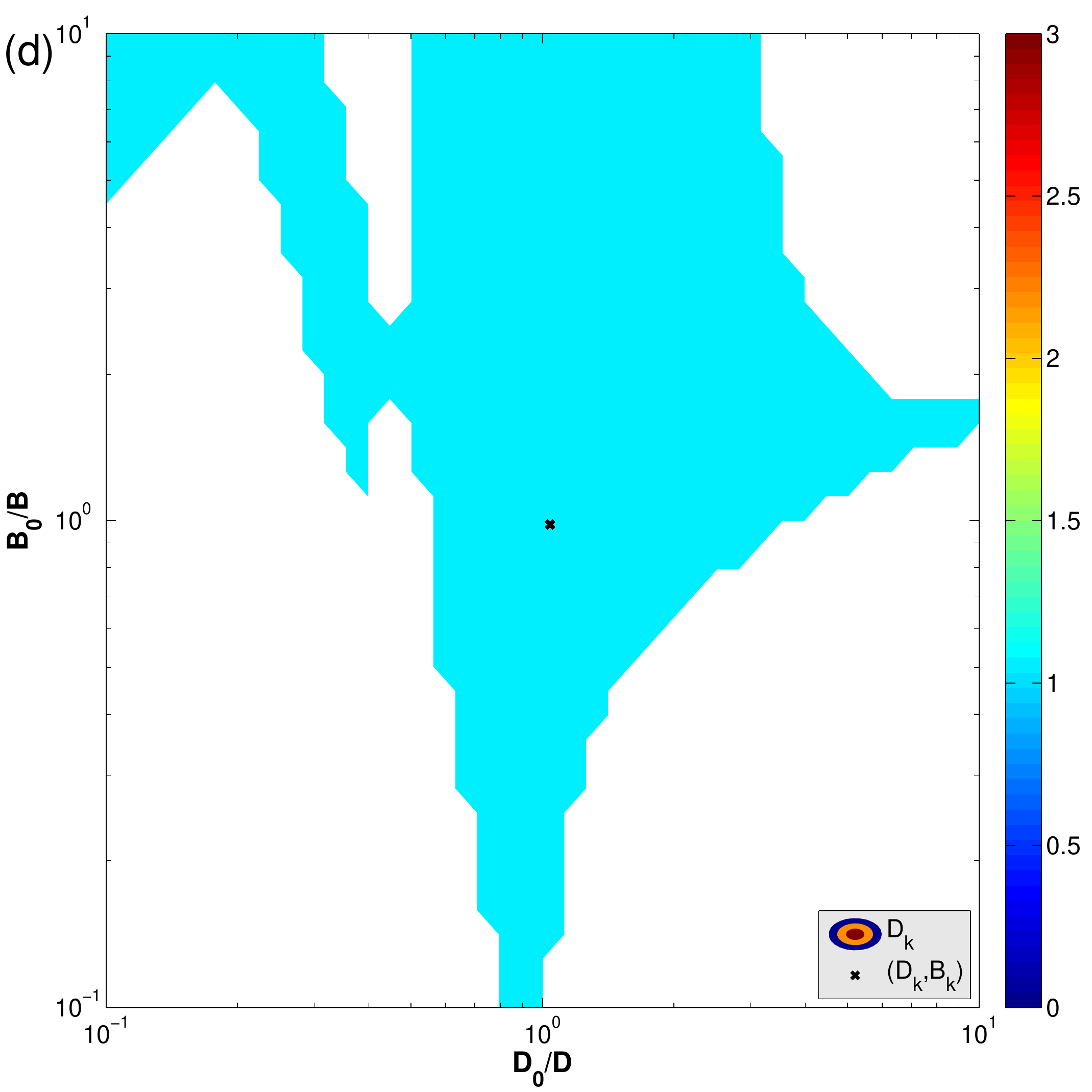}
\par\end{centering}

\protect\caption{Color plot for $D_{50}/D$ merged with convergence plot between $(D_{50}/D$,$B_{50}/B)$
for initial starting points $D_{0}/D$ and $B_{0}/B$ after 50 iterations
for: (a) suggested Kalman filter with large $\boldsymbol{R}$ and
large $\boldsymbol{Q}$, (b) Kalman filter with $p=0$, (c) Kalman
filter with $p=0.01$ , and (d) non-linear least square method. For
generated data with noises of \textpm 5\% for parameters and \textpm 10\%
for measurements.}
\label{fig: figure7}
\end{figure}

The method with the selections of $\boldsymbol{Q}$ and $\boldsymbol{R}$,
is compared with with the second, the third, and the fourth method
that are explained in subsection \autoref{sub:Selecting--and}, and
summarized in \ref{Table:1}. The $\boldsymbol{R}$ in the second
and third methods is chosen as the maximum difference between generated
data, $z$, and generated data with the noises, $h(x)$. The $\boldsymbol{P}_{0}^{-}$
for the second and third method is chosen as in Eq. (\ref{eq:Initial covaraince error}),
the same as in the suggested method. \autoref{fig: figure7}(a) shows
that 92.1\% of the initial selected $D_{0}/D$ and $B_{0}/B$ give
convergent results for diffusion constants. Around 98.3\% of those
values are converged to 1.042 after 50 iterations, while the 1.7\%
left (blue area in the color plot) almost converged to the same point
after 180 iterations. The Kalman filter with $p=0.01$ gives almost
the same convergent area of $D_{50}/D$ as the Kalman filter with
$p=0$ around 80\%, but it speeds the rate of convergence as shown
in \autoref{fig: figure7}(c), while Kalman filter with $p=0$ needs
a large number of iteration for the parameters to converge as shown
in \autoref{fig: figure7}(b). The non-linear least square speeds
the rate of convergence but with 40.93\% of the initial selected parameters
that give convergent results for diffusion constants as shown in \autoref{fig: figure7}(d). 

The suggested Kalman filter covers a wide range of the initial selected
values for $D_{0}$ and $B_{0}$ that give convergent results for
diffusion $D_{50}$ compared with Kalman filter with $p=0$, Kalman
filter with $p=0.01$, and non-linear least square. This makes it
an appropriate method to determine diffusion coefficient if the \textit{a
priori} information is rare or there is a large variation in the parameters,
such as bone as an inhomogeneous material. Also, it speeds the rate
of convergence. The Kalman filter with $p=0.01$ might be a good choice
if the \textit{a priori} information in the adjacent area of the seeking
parameters are enough. Also, the non-linear least square might be
one choice but for small range of initial selected values for $D_{0}$
and $B_{0}$. 

\begin{figure}[H]
\begin{centering}
\includegraphics[width=0.5\textwidth,height=0.5\textwidth]{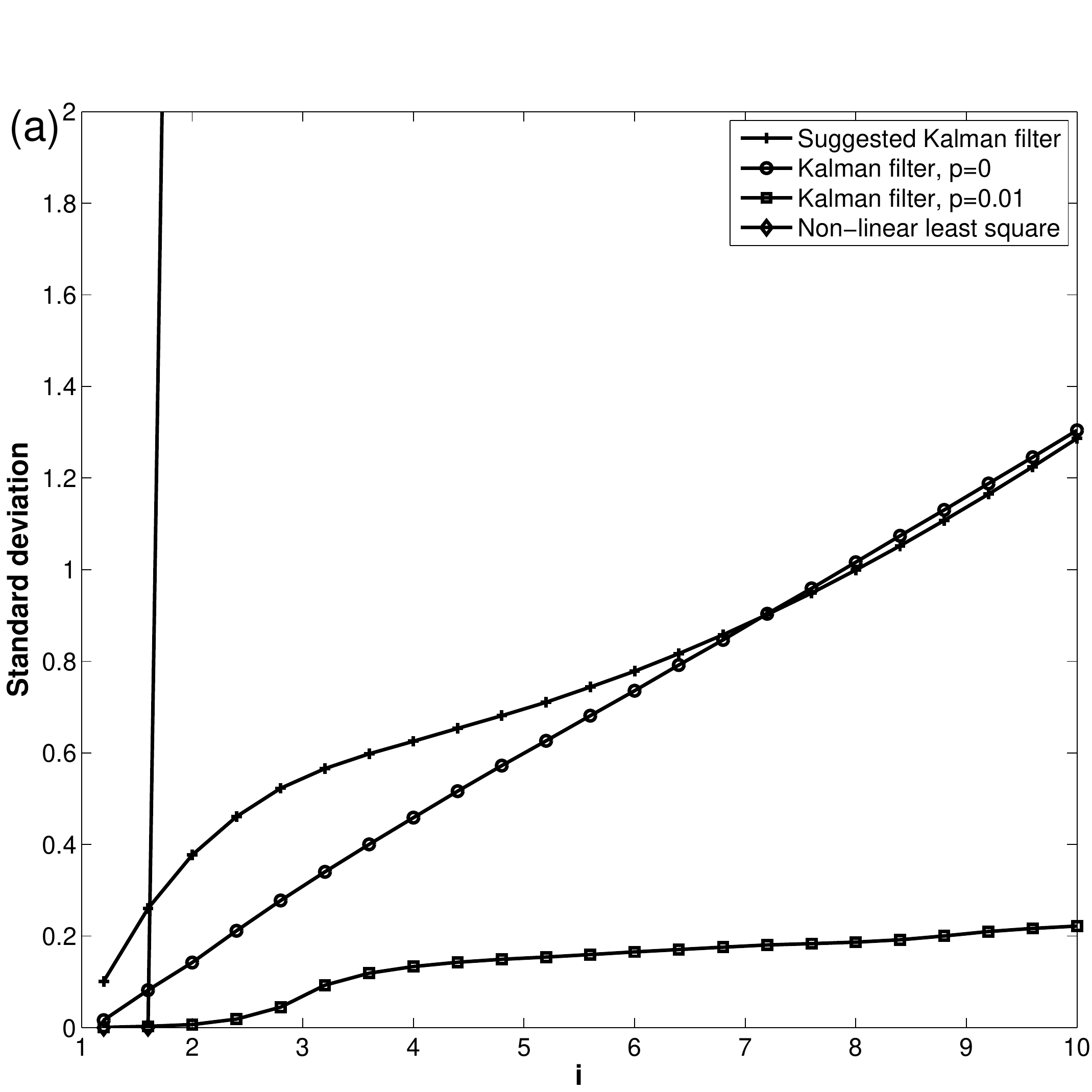}\includegraphics[width=0.5\textwidth,height=0.5\textwidth]{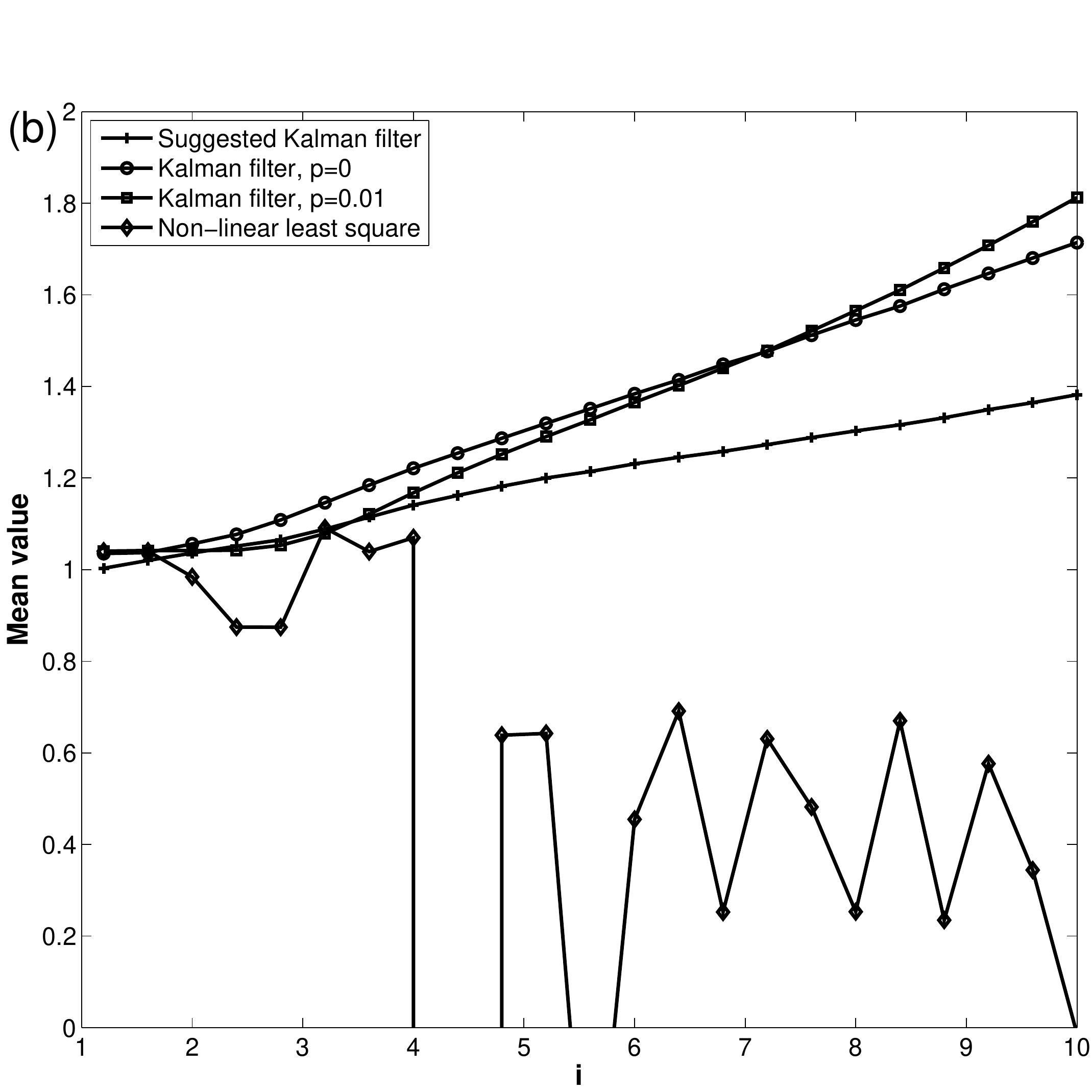}
\par\end{centering}

\protect\caption{Standard deviations and mean values for obtained $D_{5}/D$ against
ranges of selected initial parameters for the four methods after 5
iterations, (a) standard deviation and (b) mean value. For generated
data with noises of \textpm 5\% for parameters and \textpm 10\% for
measurements. }
\label{fig: figure8}
\end{figure}

The effect of different ranges for $D_{0}$ and $B_{0}$ against the
standard deviations and mean values for the four methods after 5 iterations
are shown in \autoref{fig: figure8}. The ranges are selected to start
from a small range and end with a large range as 

\[
\frac{1}{i\, D}<\text{selected initial ranges}<i\, D,
\]

\[
\frac{1}{i\, B}<\text{selected initial ranges}<i\, B,
\]
where $i=1.2,1.6,2,......,10$. The figure shows that the standard
deviations and mean values increase as $i$ increases. A reason for
that could be that the number of initial parameters increases as the
ranges increase. It can be noticed that Kalman filter with $p=0.01$
gives the smallest standard deviations while the suggested method
gives standard deviations closest to Kalman filter with $p=0$. On
the other hand, the mean values obtained by suggested method is found
to be the smallest among the other three methods. This is an indication
that the suggested method and Kalman filter with $p=0.01$ might converge
around the same speed. On the contrary, the non-linear least square
gives very large values for standard deviations for $i>1.6$, these
values are above 2 and excluded from the graph as shown in \autoref{fig: figure8}(a),
and gives large mean values for $D_{5}/D$ as shown in \autoref{fig: figure8}(b).
A possible reason for that is the divergent result that obtained after
5 iterations, it has some large negative values for $D_{5}/D$. 

\begin{figure}[H]
\begin{centering}
\includegraphics[width=0.5\textwidth,height=0.5\textwidth]{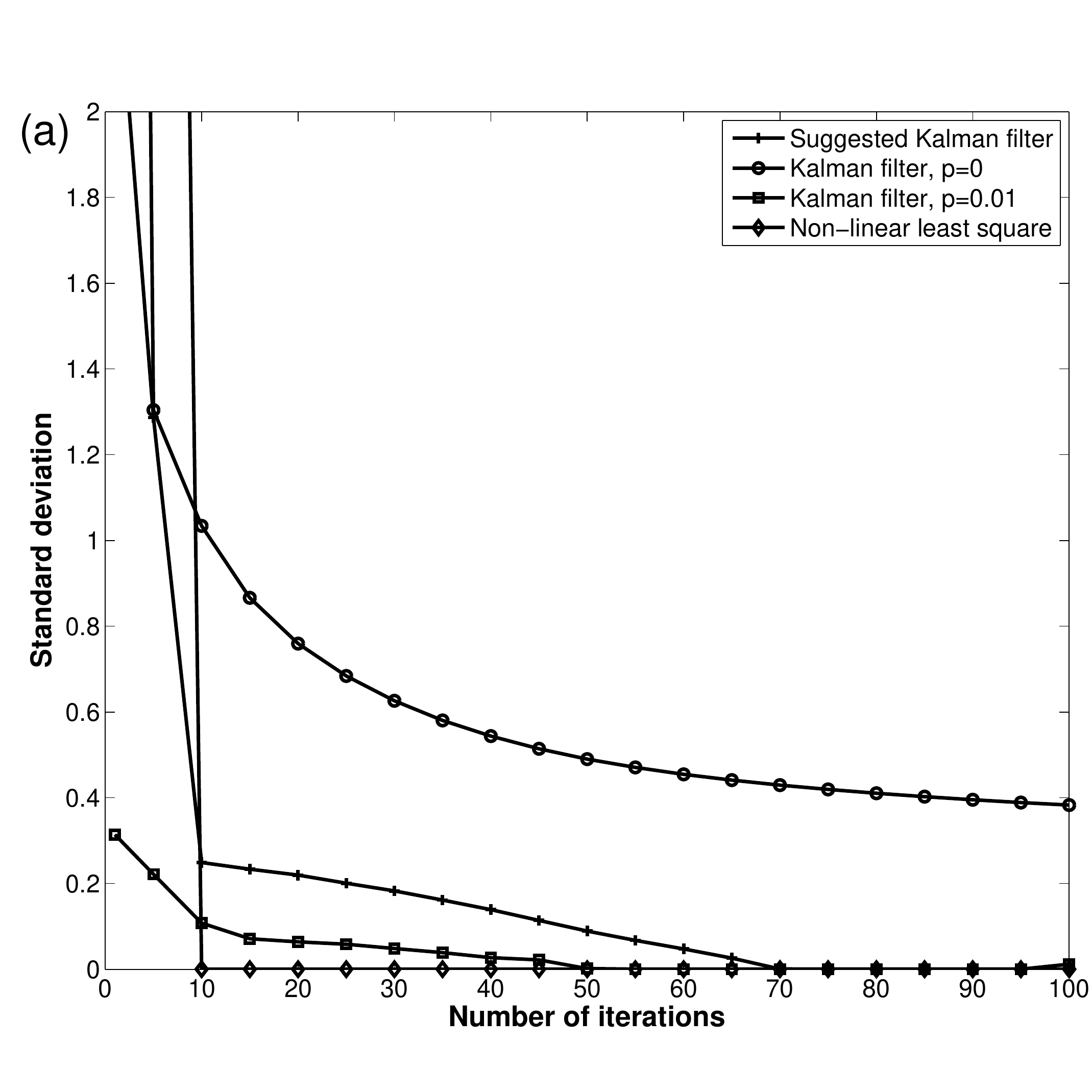}\includegraphics[width=0.5\textwidth,height=0.5\textwidth]{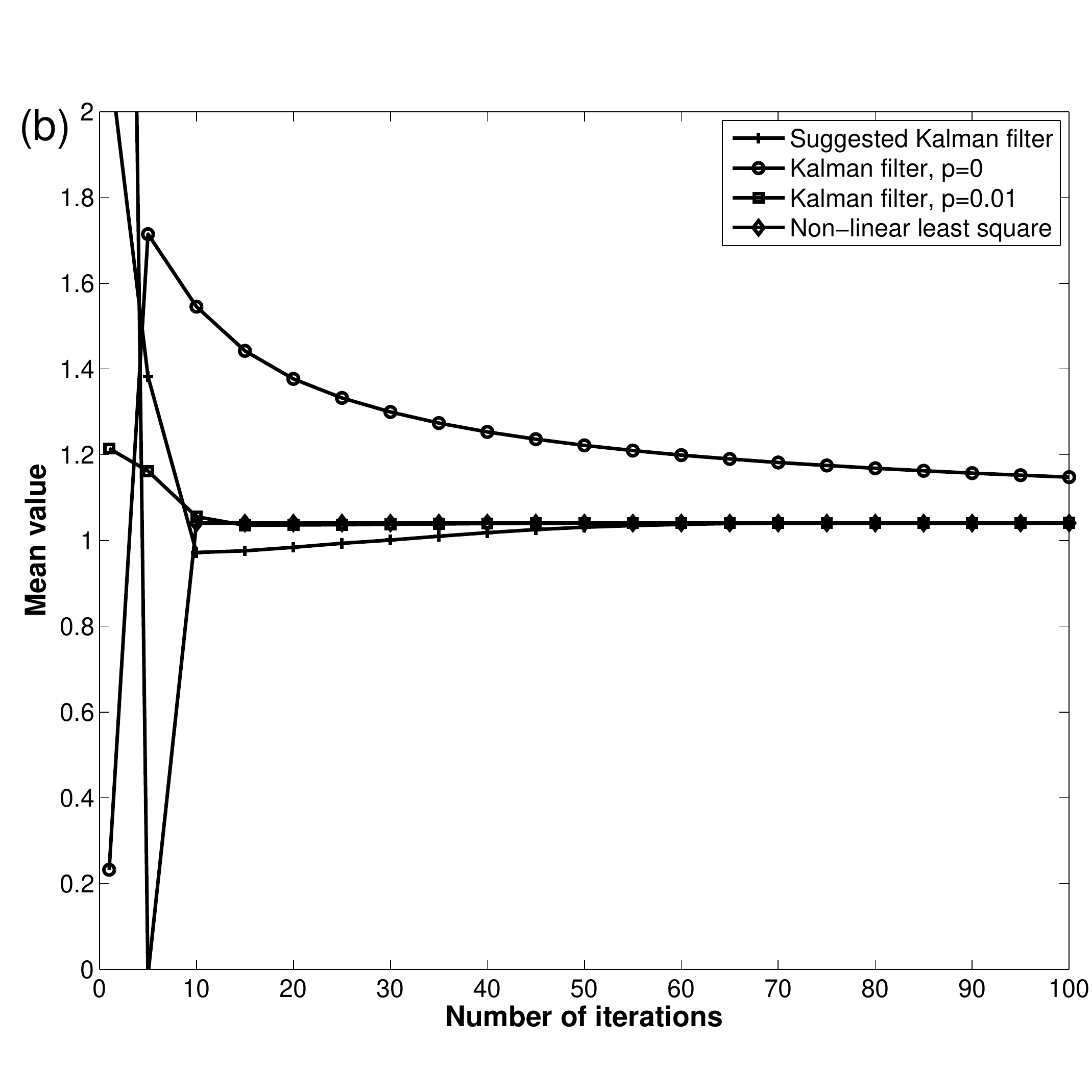}
\par\end{centering}

\protect\caption{Standard deviations and mean values for obtained $D_{k}$ against
number of iterations for the four methods after 5 iterations, (a)
standard deviation and (b) mean value. For generated data with noises
of \textpm 5\% for parameters and \textpm 10\% for measurements. (
points over 2 and less than 0 are excluded from the graph).}
\label{fig:figure9}
\end{figure}

\autoref{fig:figure9} shows the standard deviations and mean values
of the obtained $D_{k}$, $k=1,2,....,100$ against number of iterations
for large initial range, $i=10$, for the four methods. The standard
deviation values for $D_{k}$ for the suggested Kalman filter decrease
slowly up to 70 iterations, and the mean value increases as number
of iterations increases to 70 iterations, and almost stable after
that. The reason for that is some of the obtained values of $D_{k}$
were stuck to small values up to the 70 iterations, moving with small
steps of $\boldsymbol{K}_{k}$ until they reached to good predicted
values to move with larger steps. 

The standard deviation and mean value using the Kalman filter with
$p=0$ decrease slowly as the number of iterations increases. This
is expected since $\boldsymbol{R}$ in the second method was chosen
to a small value and $\boldsymbol{Q}$ vanishes, which makes the Kalman
gain is small. On the other hand, the standard deviation and mean
values of $D_{k}$ for Kalman filter with $p=0.01$ and for non-linear
least square converge quicker than the others. This is an indication
that methods 3 and 4 can be used effectivly if the \textit{apriori}
information fortunately adjacent to the seeking parameters. 

The determination of diffusion constants for generated data with \textpm 5\%
parameters noise and \textpm 50\%, and \textpm 100\% measurements
noise almost follow the same trend as in \textpm 5\% parameters noise
and \textpm 10\% measurements noise. \autoref{Table: table2} shows
the standard deviation, mean , and the percentage of numbers of convergent
$D_{50}/D$ over total number of $D_{50}/D$ (D.N\%) for the selected
range with $i=10$, for the four methods.

\begin{center}
\begin{table}[H]
\begin{centering}
\resizebox{\textwidth}{!}{
\begin{tabular}{>{\centering}m{0.25\textwidth}>{\centering}m{0.3\textwidth}>{\centering}m{0.25\textwidth}>{\centering}m{0.22\textwidth}>{\centering}m{0.1\textwidth}}
\hline 
Random noises & Method & Standard deviation & Mean  & $D.N\%$\tabularnewline
\hline 
\textpm 10\% & Suggested Kalman filter & 0.09 & 1.03 & 92.09\tabularnewline
 & Kalman filter, $p=0$ & 0.49 & 1.22 & 80.90\tabularnewline
 & Kalman filter, $p=0.01$ & 0.03 & 1.04 & 80.19\tabularnewline
 & Non-linear least square & 8.71$\times$10\textsuperscript{-09} & 1.04 & 40.93\tabularnewline
\textpm 50\% & Suggested Kalman filter & 0.06 & 0.93 & 91.96\tabularnewline
 & Kalman filter, $p=0$ & 0.54 & 1.06 & 80.37\tabularnewline
 & Kalman filter, $p=0.01$ & 0.003 & 0.94 & 80.67\tabularnewline
 & Non-linear least square & 6.19$\times$10\textsuperscript{-16} & 0.94 & 45.63\tabularnewline
\textpm 100\% & Suggested Kalman filter & 0.03 & 1.69 & 92.03\tabularnewline
 & Kalman filter, $p=0$ & 0.66 & 1.44 & 85.24\tabularnewline
 & Kalman filter, $p=0.01$ & 0.22 & 1.66 & 85.66\tabularnewline
 & Non-linear least square & 0.064 & 1.69 & 42.47\tabularnewline
\hline 
\end{tabular}}
\par\end{centering}

\protect\caption{Standard deviations, mean values and D.N\% for the obtained $D_{50}/D$
for the four methods with noises of \textpm 5\% for parameters and
\textpm 10\%, \textpm 50\%, and \textpm 100\% for measurements. }
\label{Table: table2}
\end{table}

\par\end{center}

The table shows that the suggested Kalman filter gives a compromise
results compared with the other methods. The most important difference
is the percentage of obtained $D_{50}/D$ that give convergent results
(around 92\%), that is found to be large compared with the others,
which means high possibilities to determine the diffusion coefficients
from the selected initial range even it was a large range, and for
measurements noise up to \textpm 100\%. The table also shows that
the suggested Kalman filter speeds the rate of convergence as well
as Kalman filter with $p=0.01$ and non-linear least square but with
higher D.N\%. By comparing standard deviations and mean values as
in the table, the standard deviations for Kalman filter with $p=0$
give the largest values among the other methods, and the mean values
are pretty a way from the convergent parameters. A reason for that
is Kalman filter with $p=0$ converges to many different points with
small Kalamn gain step, which means more iterations are needed for
convergence. In sum, the suggested method can be applied effectively
for both rare and sufficient information about the seeking parameters,
and can be applied for a wide range of initial parameters. 

The behavior of mean square error for a selected real bovine bone
sample is found almost the same as the generated one in \autoref{fig:figure 1}.
The initial predicted parameters $D_{0}$ and $B_{0}$ are chosen
to be $41\times41$ combinations, with $0.001<D_{0}<0.1$ mm\textsuperscript{2}/min
and $6<B_{0}<600$ \textgreek{m}S/mm, which constructs a large combinations
between the initial parameters. The initial predicted parameters error
$\boldsymbol{P}_{0}^{-}$, covariance error for the parameters $\boldsymbol{Q}$,
and covariance error for the measurements $\boldsymbol{R}$ are chosen
based on both suggested method and these combinations. \autoref{fig:figure10}
shows the convergence plot for the two parameters $D$ and $B$ that
obtained using the suggested Kalman filter for the real sample after
1, 10, 50, and 250 iterations. 

\begin{center}
\begin{figure}[H]
\begin{centering}
\includegraphics[width=0.5\textwidth,height=0.5\textwidth]{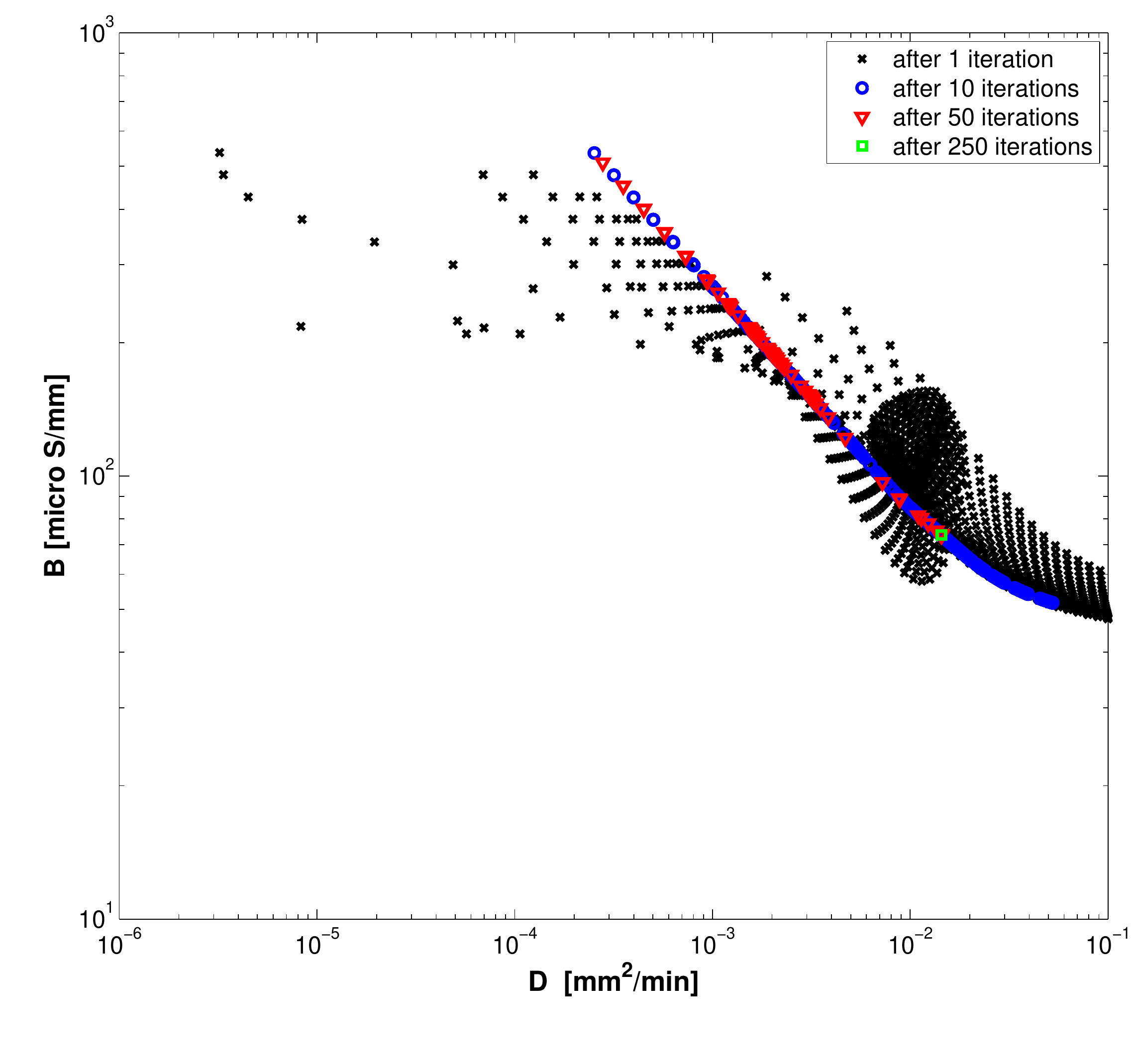}
\par\end{centering}

\protect\caption{Convergence between $D_{k}$ and $B_{k}$ obtained using suggested
method for a real sample for 1, 10, 50, and 250 iterations\label{fig:figure10}}
\end{figure}

\par\end{center}

The initial predicted parameters that give convergent results for
$D_{k}$ are found to be 92.2 \% from the combination, that converged
around a specific line after some iterations as in 10 and 50 iterations
(see \autoref{fig:figure10}). Around 92.5\% of them converged to
the seeking unknown parameters $D=0.0144$ mm\textsuperscript{2}/min
and $B=73.602$ \textgreek{m}S/mm after 50 iterations, while the 7.5\%
left converged to the same place almost after 250 iterations. Small
Kalman gain steps at the 7.5\% might be a possible reason for the
slowly convergence. The convergence line is expected since the mean
square error has smaller values along this line as shown in \autoref{fig:figure 1}. 

The conductivity versus time for analytical model Eq. (\ref{eq:Conductivity equation})
using $D$ and $B$ obtained by the suggested Kalman filter and experimental
data for the real sample is shown in \autoref{fig:figure11}. The
figure shows that the analytical function fits very well with the
experimental data accompanied with mean square error of 0.85 (\textgreek{m}S/mm)\textsuperscript{2}. 

\begin{center}
\begin{figure}[H]
\begin{centering}
\includegraphics[width=0.5\textwidth,height=0.5\textwidth]{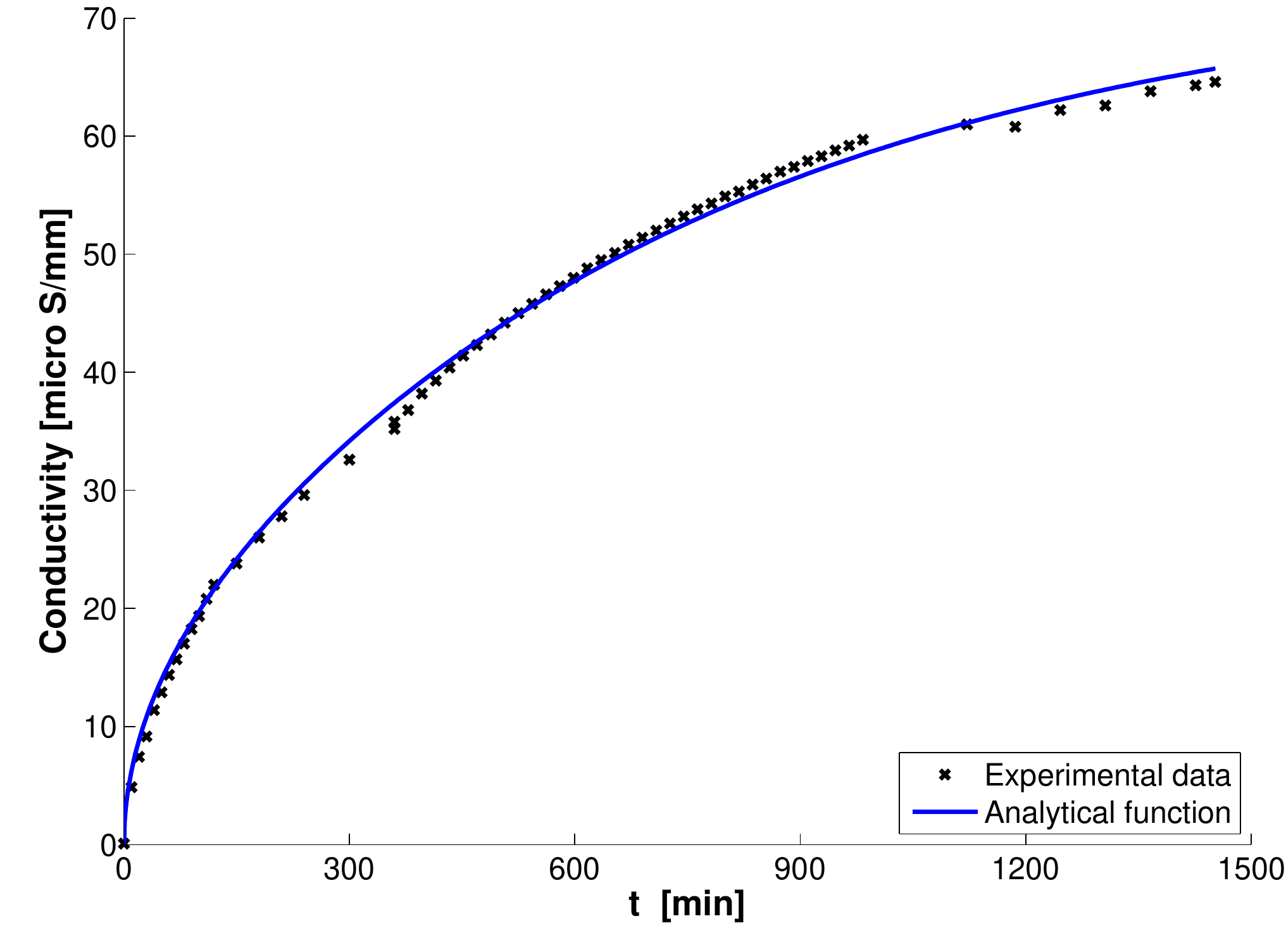}
\par\end{centering}

\protect\caption{Conductivity versus time for analytical and experimental for the real
sample\label{fig:figure11}}
\end{figure}

\par\end{center}

\section{Conclusions}

Four methods to choose appropriate initial parameters, covariance
errors for parameters, and covariance errors for measurements, required
for Kalman filter for determination of material parameters, is investigated
in this work. The methods are applied to generated data with \textpm 5\%
parameters noise and \textpm 10\%, \textpm 50\%, and \textpm 100\%
measurements noise for known parameters. The suggested method covers
a wider range of the initial suggested values for the parameters than
the standard Kalman filters, and non-linear least square, which enhances
the possibilities of convergence around the seeking parameters. The
suggested method speeds the rate of convergence compared with the
other methods. Very good results are obtained for diffusion coefficient
in bovine bone as a case study.

\appendix
\renewcommand*{\thesection}{Appendix A}

\section{Non-linear least squares \label{sec:Appendix A}}

\renewcommand{\theequation}{A-\arabic{equation}}
\setcounter{equation}{0}  

The sum of the squared residuals is written

\begin{equation}
\Phi=(z-h)^{\text{T}}(z-h)
\end{equation}
By putting the derivative of $\Phi$\nomenclature{$\Phi$}{sum of the squared residuals between $z$ and $h$}
with respect to $x$ to zero, a set of $n$ equations is obtained, 

\begin{equation}
\frac{\partial\Phi}{\partial x}\mbox{=}0\ \ \mbox{\ensuremath{\Rightarrow}}\ \mathbf{\ H}(x)^{\text{T}}\{z-h(x)\}=0\,.\label{eq:derivativePhi-1}
\end{equation}
Insertion of Eq. (\ref{eq:Taylor}) into Eq. (\ref{eq:derivativePhi-1})
gives the following,

\begin{equation}
\mathbf{H}^{\text{T}}\{z-h(\hat{x}_{k})\}\approx\mathbf{\mathbf{H}^{\text{T}}H}(x-\hat{x}_{k})\,.
\end{equation}
Provided that the $n\times n$ matrix $\mathbf{\mathbf{H}^{\text{T}}H}$
is non-singular, one obtains the following approximation,

\begin{equation}
x\approx\hat{x}_{k}+(\mathbf{\mathbf{H}^{\text{T}}H})^{-1}\mathbf{H}^{\text{T}}\{z-h(\hat{x}_{k})\}.\label{eq:least square-1}
\end{equation}
For linear systems the Jacobian $\mathbf{H}$ is independent of $x$,
which makes the solution in Eq. (\ref{eq:least square-1}) exact. 

\renewcommand*{\thesection}{Appendix B}

\section{Kalman filter derivation\label{sec:Appendix-B}}

\renewcommand{\theequation}{B-\arabic{equation}}
\setcounter{equation}{0}  

The variance of the errors of the parameters before, $\boldsymbol{P}_{k}^{-}$,
and after, $\boldsymbol{P}_{k}$, the iterative update are

\begin{equation}
\boldsymbol{P}_{k}^{-}=\text{Var\ensuremath{(x_{k}-\hat{x}_{k-1})}},\ \ \text{and}\ \ \boldsymbol{P}_{k}\mbox{=}\text{Var}(x_{k}-\hat{x}_{k})\,.\label{eq:corrected parameters error equation-1-1}
\end{equation}
The function $\text{Var}(a)$ is an $l\times l$ matrix with the elements
$\alpha_{ij}$ given by the vector $a=(a_{1},...a_{l})^{\text{T}}$
as follows (cf. \citep{Courant(1953)})

\begin{equation}
\alpha_{ij}=\text{E}(x_{i}x_{j})-\text{E}(x_{i})\text{E}(x_{j})\,,
\end{equation}
where $E(\vartheta)$ is the statistical mean value of the stochastic
variable $\vartheta$. Hence, the matrices $\boldsymbol{P}_{k}$ and
$\boldsymbol{P}_{k-1}^{-}$ are symmetric with the dimension $n\times n$.

Using Eqs. (\ref{eq:corrected parameters error equation-1-1}) and
(\ref{eq:matparameter noise}), a relation between $\boldsymbol{P}_{k}^{-}$
and $\boldsymbol{P}_{k}$ is obtained as,

\begin{equation}
\boldsymbol{P}_{k+1}^{-}=\text{Var}(x_{k+1}-\hat{x}_{k})=\text{Var}(x+w_{k+1}-\hat{x}_{k})=\text{Var}(x_{k}-w_{k}+w_{k+1}-\hat{x}_{k})=\text{Var}(x_{k}-\hat{x}_{k})+\text{Var}(-w_{k}+w_{k+1})=\text{Var}(x_{k}-\hat{x}_{k})+\text{2Var}(w_{k})=\boldsymbol{P}_{k}+\boldsymbol{Q}_{k}\label{eq:def_of_Q_k-1}
\end{equation}
where $\boldsymbol{Q}_{k}=2\text{Var}(w_{k})$ $\boldsymbol{Q}_{k}$\nomenclature{$\boldsymbol{Q}_{k}$, $\boldsymbol{Q}$}{covariance error of the parameters at iteration  $k$, constant covariance}
is an $n\times n$ matrix. Note that the covariance of the supposedly
uncorrelated quantities $(x-\hat{x}_{k})$ and $w_{k}$ respectively
vanishes.

On the other hand, substituting $\hat{x}_{k}$ in Eq. (\ref{eq:corrected parameters error equation-1-1})
by using Eq. (\ref{eq:kalman}), $\boldsymbol{P}_{k}$ can be expressed
as

\[
\boldsymbol{P}_{k}=\text{Var}[x_{k}-(\hat{x}_{k-1}+\mathbf{K}{}_{k}\{z_{k}-h(\hat{x}_{k-1})\})]\,.
\]
Replacing $z_{k}$ according to Eq. (\ref{eq:model-properties}) leads
to

\begin{equation}
\boldsymbol{P}_{k}=\text{Var}(x_{k}-\hat{x}_{k-1}-\mathbf{K}{}_{k}\{h(x_{k})+v_{k}-h(\hat{x}_{k-1})\})=\text{Var}(x_{k}-\hat{x}_{k-1})-\mathbf{K}{}_{k}\text{Cov(}h(x_{k})-h(\hat{x}_{k-1}),x_{k}-\hat{x}_{k-1})-\text{Cov}(x_{k}-\hat{x}_{k-1},h(x_{k})-h(\hat{x}_{k-1}))\mathbf{K}{}_{k}^{\text{T}}+\mathbf{K}{}_{k}\text{Var}(h(x_{k})-h(\hat{x}_{k-1}))\mathbf{K}{}_{k}^{\text{T}}+\mathbf{K}{}_{k}\text{Var}(v_{k})\mathbf{K}{}_{k}^{\text{T}}\,,
\end{equation}
where the function $\text{Cov}(a,b)$ gives the covariance of the
stochastic vectors $a$\nomenclature{$a$, $b$}{two stochastic vectors}
and $b$. The function $\text{Cov}(a,b)$ is an $m\times m$ matrix
with the elements $\alpha_{ij}$\nomenclature{$\alpha_{ij}$}{elements of  covariance matrix of stochastic vectors $a$, and $b$.}
given by the stochastic variables $a=(a_{1},...a_{m})^{\text{T}}$
and $b=(b_{1},...b_{m})^{\text{T}}$ as follows (cf. \citep{Courant(1953)}),

\begin{equation}
\alpha_{ij}=\text{E}(a_{i}b_{j})-\text{E}(a_{i})\text{E}(b_{j})\,.
\end{equation}

It is used that the covariance between the elements of $h(x_{i})$
and $v_{k}$ for any $i$ and $k$ vanishes. The Taylor series in
Eq. (\ref{eq:Taylor}) giving $h(x_{k})-h(\hat{x}_{k-1})=\mathbf{H}_{k}(x_{k}-\hat{x}_{k-1})$
results in

\begin{equation}
\boldsymbol{P}_{k}=\boldsymbol{P}_{k}^{-}-\mathbf{K}{}_{k}\mathbf{H}_{k}\boldsymbol{P}_{k}^{-}-\boldsymbol{P}_{k}^{-}\mathbf{H}_{k}^{\text{T}}\mathbf{K}_{k}^{\text{T}}+\mathbf{K}{}_{k}\mathbf{H}_{k}\boldsymbol{P}_{k}^{-}\mathbf{H}_{k}^{\text{T}}\mathbf{K}_{k}^{\text{T}}+\mathbf{K}{}_{k}\boldsymbol{R}_{k}\mathbf{K}{}_{k}^{\text{T}}=(\boldsymbol{I}-\mathbf{K}_{k}\mathbf{H}_{k})\boldsymbol{P}_{k}^{-}(\boldsymbol{I}-\mathbf{K}_{k}\mathbf{H}_{k})^{\text{T}}+\mathbf{K}{}_{k}\boldsymbol{R}_{k}\mathbf{K}{}_{k}^{\text{T}}\,,\label{eq:corrected parameter error total-1-1}
\end{equation}
where $R_{k}$ is the covariance error of the measurements and an
$N\times N$ matrix defined as follows

\begin{equation}
\boldsymbol{R}_{k}=\text{Var}(v_{k})\,.\label{eq:Covariance error for measurements-1}
\end{equation}

Since the $x_{k}$ and $v_{k}$ are mutually uncorrelated, $\boldsymbol{P}_{k}$
becomes a diagonal matrix that contains errors between the parameters
before and after an iteration, the $K_{k}$ that minimizes the error
can be obtained by taking the derivative of the trace $\text{Tr}(\boldsymbol{P}_{k})$
with respect to $\mathbf{K}_{k}$ and putting it equal to zero. Taking
the trace of the first equality in Eq. (\ref{eq:corrected parameter error total-1-1})
provides

\begin{equation}
\text{Tr}(\boldsymbol{P}_{k})=\text{Tr}(\boldsymbol{P}_{k}^{-})-\text{Tr}(\mathbf{K}{}_{k}\mathbf{H}_{k}\boldsymbol{P}_{k}^{-})-\text{Tr}(\boldsymbol{P}_{k}^{-}\mathbf{H}_{k}^{\text{T}}\mathbf{K}_{k}^{\text{T}})+\text{Tr}(\mathbf{K}{}_{k}\mathbf{H}_{k}\boldsymbol{P}_{k}^{-}\mathbf{H}_{k}^{\text{T}}\mathbf{K}_{k}^{\text{T}})+\text{Tr}(\mathbf{K}{}_{k}\boldsymbol{R}_{k}\mathbf{K}{}_{k}^{\text{T}})\label{eq:Trace of P-1}
\end{equation}

Using following identities for the matrix trace (cf. \citep{Petersen(2008)}) 

\begin{equation}
\text{Tr}(\boldsymbol{P}_{k}^{-}\mathbf{H}{}_{k}^{\text{T}}\mathbf{K}{}_{k}^{\text{T}})=\text{Tr}(\mathbf{K}{}_{k}\mathbf{H}_{k}\boldsymbol{P}_{k}^{-})\,,\ \frac{\partial\text{Tr}(\boldsymbol{AC})}{\partial\boldsymbol{A}}\mbox{=}\boldsymbol{C}^{T}\,\text{and}\,\frac{\partial\text{Tr}(\boldsymbol{AF}\boldsymbol{A}^{T})}{\partial\boldsymbol{A}}\mbox{=}2\boldsymbol{AF}\,.
\end{equation}
The last equality requires that the matrix $\boldsymbol{F}$ is symmetric.
The derivative of Eq. (\ref{eq:Trace of P-1}) with respect to $\mathbf{K}{}_{k}$
can be written

\begin{equation}
\frac{\text{\ensuremath{\partial}Tr}(\boldsymbol{P}_{k})}{\partial\mathbf{K}_{k}}=-2\boldsymbol{P}_{k}^{-}\mathbf{H}_{k}^{\text{T}}+2\mathbf{K}{}_{k}\mathbf{H}_{k}\boldsymbol{P}_{k}^{-}\mathbf{H}_{k}^{\text{T}}+2\mathbf{K}{}_{k}\boldsymbol{R}_{k}\label{eq:derivative of trace-1}
\end{equation}

The optimal $\mathbf{K}{}_{k}$ can be obtained by putting the right
term in Eq. {[}\ref{eq:derivative of trace-1}{]} to zero as next

\begin{equation}
\mathbf{K}{}_{k}=\boldsymbol{P}_{k}^{-}\mathbf{H}_{k}^{\text{T}}(\mathbf{H}_{k}\boldsymbol{P}_{k}^{-}\mathbf{H}_{k}^{\text{T}}+\boldsymbol{R}_{k})^{-1}\,.\label{eq:Kalman gain-1}
\end{equation}

The $\boldsymbol{P}_{k}$ associated to the optimal $\mathbf{K}_{k}$
can be written as in Eq. {[}\ref{eq:Covariance error P-1}{]} by substituting
Eq. {[}\ref{eq:Kalman gain-1}{]} into Eq. (\ref{eq:corrected parameter error total-1-1})
as next 

\[
\boldsymbol{P}_{k}=(\boldsymbol{I}-\mathbf{K}{}_{k}\mathbf{H}_{k})\boldsymbol{P}_{k}^{-}+\{-\boldsymbol{P}_{k}^{-}\mathbf{H}_{k}^{\text{T}}+\mathbf{K}{}_{k}(\mathbf{H}_{k}\boldsymbol{P}_{k}^{-}\mathbf{H}_{k}^{\text{T}}+\boldsymbol{R}_{k})\}\mathbf{K}{}_{k}^{\text{T}}
\]

Using Eq. (\ref{eq:Kalman gain-1}) readily gives

\begin{equation}
\boldsymbol{P}_{k}=(\boldsymbol{I}-\mathbf{K}{}_{k}\mathbf{H}_{k})\boldsymbol{P}_{k}^{-}\,,\label{eq:Covariance error P-1}
\end{equation}
and consequently

\begin{equation}
\boldsymbol{P}_{k+1}^{-}=(\boldsymbol{I}-\mathbf{K}{}_{k}\mathbf{H}_{k})\boldsymbol{P}_{k}^{-}+\boldsymbol{Q}_{k}\,,\label{eq:P-recursion-1}
\end{equation}

\newcommand{\noop}[1]{} 

\section*{\refname}

{\footnotesize{}\bibliographystyle{elsarticle-num}
\bibliography{ReferencesMethodology}
}
\end{document}